\documentclass[11pt,a4paper]{article}

\usepackage{graphicx}
\usepackage{caption}
\usepackage{subcaption}
\usepackage{float}
\usepackage{siunitx}
\usepackage{amsmath}
\usepackage{hyperref}
\usepackage{multirow}
\usepackage{booktabs}
\usepackage[numbers]{natbib}
\usepackage{geometry}
\graphicspath{ {img/} }

\begin{document}

\begin{center}
{\LARGE \textbf{Temperature and Pressure Dependent Vibrational Properties of Pristine and Doped Vacancy-Ordered Double Perovskite}}

\vspace{1.5cm}

{\large Aalok Tiwari,$^{1}$ Karamjyoti Panigrahi,$^{2}$ Mrinmay Sahu,$^{1}$}

{\large Sayan Bhattacharyya,$^{2}$ and Goutam Dev Mukherjee$^{1}$}

\vspace{1cm}

{\itshape $^{1}$National Centre for High Pressure Studies, Department of Physical Sciences,}

{\itshape Indian Institute of Science Education and Research Kolkata,}

{\itshape Mohanpur Campus, Mohanpur 741246, Nadia, West Bengal, India}

\vspace{0.5cm}

{\itshape $^{2}$Department of Chemical Sciences, and Centre for Advanced Functional Materials,}

{\itshape Indian Institute of Science Education and Research (IISER) Kolkata,}

{\itshape Mohanpur-741246, India}

\vspace{1cm}

(Dated: \today)
\end{center}

\vspace{0.5cm}

\begin{abstract}
Understanding lattice dynamics and structural transitions in vacancy-ordered double perovskites is crucial for developing lead-free optoelectronic materials, yet the role of dopants in modulatingthese properties remains poorly understood. We investigate the vibrational and optical properties of pristine and Antimony(Sb)-doped Cs$_2$TiCl$_6$ vacancy-ordered double perovskite through temperature-dependent Raman spectroscopy (4--273 K), high-pressure studies (0-- \~30 GPa), ambient powder XRD, and photoluminescence measurements. Sb doping improves phase purity, reducing impurity-related Raman modes present in pristine samples. Most notably, Sb-doped samples exhibit an anomalous Raman mode M$_1$ appearing exclusively below 100 K at 314--319 cm$^{-1}$, accompanied by changes in the temperature coefficient $\chi$ and anharmonic constant $A$ across this threshold. This behavior is absent in pristine Cs$_2$TiCl$_6$. While these observations suggest possible structural changes at low temperature, the origin of the M$_1$ mode remains unclear and may arise from disorder-activated vibrations, symmetry breaking, or dopant-induced local distortions. Low-temperature structural characterization is needed to confirm the nature of this transition. Photoluminescence shows broad self-trapped exciton emission at 448 nm with broader FWHM in Sb-doped samples (164.73 nm) compared to Bi-doped samples (138.2 nm), consistent with enhanced structural disorder. High-pressure Raman measurements reveal continuous mode hardening to 30 GPa with no phase transitions. These results demonstrate that Sb doping modulates the vibrational properties of Cs$_2$TiCl$_6$, though further investigation is required to establish the underlying mechanisms.
\end{abstract}


\section{Introduction}

Perovskites are promising materials for energy storage and optoelectronic devices such as solar cells, light-emitting diodes, and memory devices due to their excellent catalytic and photoelectric properties. Materials with the ABX$_3$ structure are collectively known as perovskite materials \cite{one}. In 3D metal halide perovskites, A is an organic or inorganic short-chain cation (MA$^+$, FA$^+$, or Cs$^+$), B is a divalent metal cation (Pb$^{2+}$, Sn$^{2+}$, or Ti$^{2+}$), and X is a halogen anion (I$^-$, Br$^-$, or Cl$^-$). The [BX$_6$]$^{4-}$ octahedra connect through corner-sharing to form an isotropic inorganic framework, with B-site cations at octahedral centers, A-site cations filling voids between the inorganic frameworks, and highly electronegative X-site halogens at octahedral corners.

\begin{figure}[htbp]
    \centering
    \includegraphics[width=0.75\linewidth]{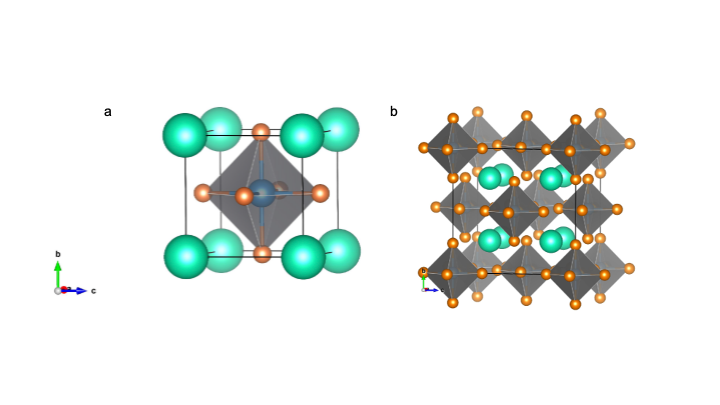}
\caption{General structure of a. $ABX_3$ and $A_2BX_6$ halide perovskite} 
    \label{Fig1}
\end{figure}

The 3D perovskite structure with corner-sharing six-coordinated octahedra provides broad conduction and valence bands that promote charge transport \cite{3ds}. 2D halide perovskites are obtained by cleaving the 3D structure along (100), (110), and (111) crystallographic planes to form sheets that stack into bulk crystalline materials as shown in Figure \ref{Fig1} \cite{lds}.

Different perovskite structures can be obtained by partially or fully substituting A and B site cations. The aristotype perovskite belongs to the Pm$\Bar{3}$m cubic space group. Distorted octahedra, lattice distortions, vacancies, ordered cations, or organic cations/inorganic clusters lead to reduced symmetry in many perovskites \cite{akke}.

Vacancy-ordered double perovskite halides have the form A$_2$BX$_6$, where one B-site cation is replaced by a vacancy \cite{P2}. These halides are crystallographically analogous to ordered perovskites with space group \textit{Fm$\bar{3}$m}. Typical examples include Cs$_2$SnI$_6$ and Cs$_2$TiBr$_6$. Systems using trivalent metals (Sb$^{3+}$, Bi$^{3+}$) have B sites occupied by metal and vacancy in a 2:1 ratio, yielding A$_3$B$_2$X$_9$ stoichiometry. Vacancies ordered along [111] planes create a two-dimensional layer of BX$_6$ octahedra \cite{akke}. High vacancy concentrations result in low conductivity and limited applications \cite{maugh}.

The optoelectronic and photophysical properties of inorganic halide perovskites (CsPbX$_3$) are primarily influenced by their soft, dynamic 3D ionic lattice. The corner-sharing [PbX$_6$]$^{4-}$ octahedral network affects electronic and phonon dispersions, modulating the dielectric environment and giving rise to optoelectronic phenomena. Understanding how octahedral networks influence lattice dynamics and anharmonicity is essential. Recent studies by Calzolari \cite{Calzolari} and Steele et al. \cite{Steele} show that symmetry-allowed optical phonon modes significantly contribute to dielectric modulation. Raman spectroscopy is an excellent tool for probing perovskite crystal dynamics, lattice anharmonicity, and electron-phonon coupling \cite{P2, motiv}. However, phonon measurements of archetypal CsPbX$_3$ perovskites cannot be quantified by Raman spectroscopy because: (i) local symmetry breaking renders intrinsic Raman modes inactive \cite{jpa}, and (ii) ultra-low frequency modes are thermally smeared into anharmonic spectral features \cite{prl}.

Investigating the atomic structure and fluctuations of octahedral building blocks in perovskites, especially mixed-halide systems, is crucial for understanding their electronic, vibrational, and structural characteristics. The fundamental octahedral building block can be better studied in lead-free inorganic vacancy-ordered double perovskites (A$_2$BX$_6$) \cite{motiv}. In A$_2$BX$_6$ systems, systematic B-site vacancies create isolated [BX$_6$]$^{2-}$ octahedral units stabilized by close-packing of X-site anions and A-site cations bridging the 12-coordinated cuboctahedral voids. The zero-dimensional [BX$_6$]$^{2-}$ network possesses molecule-like electronic features, with isolated octahedra serving as both electronic and vibrational centers, enabling direct investigation of the fundamental perovskite building block \cite{arun, berg}.

We selected Cs$_2$TiCl$_6$ because it can be synthesized via low-cost growth \cite{growth} and has very high power conversion efficiency \cite{pce}. Its indirect bandgap of 2.8 eV and photoluminescence peaks at 535–670 nm make it suitable for thin-film photovoltaic applications \cite{app}. The high Goldschmidt tolerance factor of 0.95 provides structural stability, making it thermodynamically stable and promising for solar cells. The material is lead-free, eco-friendly, and highly susceptible to temperature and pressure due to its soft metal lattice. External pressure can alter the inorganic octahedra, potentially yielding interesting photoluminescence observations and providing fundamental understanding of vacancy-ordered double perovskite halides.

\section{Experimental Methods}

Cs$_2$TiCl$_6$ was synthesized using an acid-assisted precipitation method. Two millimoles of CsCl were dissolved in 2 mL of HCl and stirred at 60$^{\circ}$C for 5 hours. After obtaining a clear solution, one millimole of TiCl$_4$ was added at 60$^{\circ}$C and stirred until cooled to room temperature. The chemical reaction follows:
\begin{center}
    2CsCl + TiCl$_4$ $\xrightarrow[]{}$ Cs$_2$TiCl$_6$
\end{center}
The resulting solution was washed with isopropanol (IPA) and centrifuged at 7000 rpm. The precipitate was dried at 60$^{\circ}$C to obtain polycrystalline Cs$_2$TiCl$_6$. HCl serves as both solvent and halide source, ensuring the desired perovskite composition.

For Sb/Bi-doped samples, 0.97 mmol of TiCl$_4$ was mixed with either 0.17 mmol of SbCl$_3$ or 0.19 mmol of BiCl$_3$ in 1 mL HCl and stirred for 15 minutes. This solution was then added to 2 mmol of CsCl dissolved in 2 mL HCl at 60$^{\circ}$C. The mixture was stirred and cooled to room temperature, followed by the same washing, centrifugation, and drying procedure to obtain Cs$_2$Ti$_{1-x}$(Sb/Bi)$_{x}$Cl$_6$ with 2\% or 3\% doping. All chemicals were purchased from Sigma-Aldrich.

Powder X-ray diffraction (XRD) was performed using a MiniFlex diffractometer (RIGAKU) with Cu K$\alpha$ radiation ($\lambda$ = 1.5418 \si{\angstrom}) over a 2$\theta$ range of \ang{5} to \ang{70}. Photoluminescence (PL) measurements were conducted using an FLS 1000 spectrometer (Edinburgh) equipped with a 450 W Xenon lamp. Raman spectroscopy and Fourier-transform infrared spectroscopy (FTIR) measurements were performed on Cs$_2$TiCl$_6$ and Cs$_2$Ti$_{1-x}$(Sb/Bi)$_{x}$Cl$_6$ samples with x = 2\% and 3\% incorporation and doping respectively.

Stability tests were performed to assess moisture and heat resistance. For water stability, 1 mL of water was added to 0.5 mg powder samples, and Raman spectra were collected at regular intervals. The Cs$_2$TiCl$_6$ (3\% Sb) system showed no spectral changes, with main T$_{2g}$ and A$_{1g}$ modes at 180 and 318 cm$^{-1}$ remaining intact. Low-intensity modes at 120 and 268 cm$^{-1}$ disappeared immediately in Cs$_2$TiCl$_6$ and after 15 minutes in Cs$_2$TiCl$_6$ (3\% Bi), likely due to impurity-water interaction. For heat stability, samples were continuously heated at 120$^{\circ}$C in a vacuum furnace. Raman spectra showed no changes after 7 hours, confirming thermal stability up to 120$^{\circ}$C.

\section{Results and Discussion}

\subsection{Structural and Vibrational Characterization}

\begin{figure}[htbp]
    \centering
    \includegraphics[width=1\linewidth]{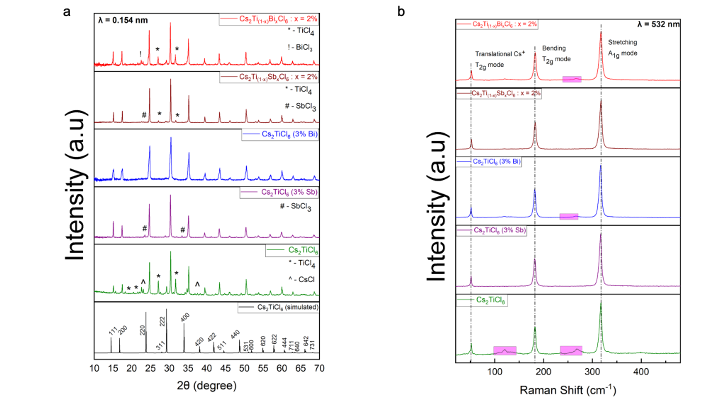}
    \caption{a. Ambient X-ray diffraction measurement using 0.154 nm X-rays. b. Ambient Raman Spectra using 532 nm laser for all synthesized samples}
    \label{Fig2}
\end{figure}

Figure \ref{Fig2}a shows the powder XRD patterns of Cs$_2$TiCl$_6$ and Cs$_2$Ti$_{1-x}$(Sb/Bi)$_{x}$Cl$_6$ (x = 2\%, 3\%) samples compared with simulated data from the Materials Project (mp-27828). The experimental peaks are shifted approximately \ang{1} to higher 2$\theta$ values relative to the simulated pattern. All samples exhibit additional peaks corresponding to unreacted starting materials: TiCl$_4$ (marked by *, mp-30092) at \ang{27.01} and \ang{31.58}, CsCl (marked by \^{}, mp-573697) at \ang{23.14} and \ang{37.11}, SbCl$_3$ (marked by \#, mp-22872) at \ang{23.43} and \ang{33.67}, and BiCl$_3$ (marked by !, mp-22908) at \ang{22.32}. 

The presence of impurity phases in all samples complicates direct comparison between pristine and doped materials posing an important limitation. In pristine Cs$_2$TiCl$_6$, impurity peaks from both TiCl$_4$ and CsCl are observed. The Cs$_2$TiCl$_6$ (3\% Sb) sample shows additional SbCl$_3$ peaks, while Cs$_2$TiCl$_6$ (3\% Bi) exhibits a missing peak at \ang{15.13}. The  Sb-doped samples show reduced impurity peak intensity compared to pristine Cs$_2$TiCl$_6$, suggesting improved phase purity with dopant incorporation. 
Room temperature Raman spectra collected using 532 nm excitation is shown in Figure \ref{Fig2}b and the observed Raman modes are summarized in Table \ref{tab:raman_modes}. Three main phonon modes are consistently observed: translational Cs$^+$ T$_{2g}$ at $\sim$50-52 cm$^{-1}$, bending T$_{2g}$ at $\sim$180-182 cm$^{-1}$, and stretching A$_{1g}$ at $\sim$312-318 cm$^{-1}$, corresponding to vibrations within the [TiCl$_6$]$^{2-}$ octahedra. Pristine Cs$_2$TiCl$_6$ and Bi-doped samples exhibit additional unidentified peaks at 120, 256, and 270 cm$^{-1}$, likely originating from the impurity phases identified in XRD. Both Sb-doped samples show cleaner spectra with no additional peaks beyond the three main modes. The Sb-doped samples also display slight frequency shifts: the A$_{1g}$ mode shifts from 318 to 312 cm$^{-1}$ in Cs$_2$TiCl$_6$ (2\% Sb), while the T$_{2g}$ bending mode shifts to 179-180 cm$^{-1}$.

\begin{table}[h]
\centering 
\caption{Raman modes (cm$^{-1}$) observed at ambient conditions}
\label{tab:raman_modes}
\begin{tabular}{lcccc}
\hline\hline
Material & T$_{2g}$ (Cs$^+$) & T$_{2g}$ (bend) & A$_{1g}$ (stretch) & Unidentified \\
\hline
Cs$_2$TiCl$_6$ & 52 & 182 & 318 & 120, 256, 270 \\
Cs$_2$TiCl$_6$ (3\% Bi) & 52 & 182 & 318 & 120, 270 \\
Cs$_2$TiCl$_6$ (3\% Sb) & 51 & 180 & 318 & --- \\
Cs$_2$TiCl$_6$ (2\% Bi) & 52 & 182 & 318 & 120, 270 \\
Cs$_2$TiCl$_6$ (2\% Sb) & 50 & 179 & 312 & --- \\
\hline\hline
\end{tabular}
\end{table}

\subsection{Photoluminescence}

\begin{figure}[htbp]
    \centering
    \includegraphics[width=1\linewidth]{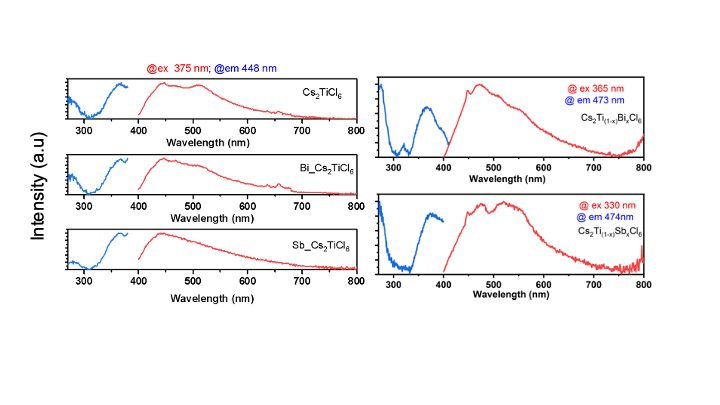}
    \caption{Excitation and emission spectra of Cs$_2$TiCl$_6$ and doped systems, including Cs$_2$Ti$_{(1-x)}$Bi$_x$Cl$_6$ and Cs$_2$Ti$_{(1-x)}$Sb$_x$Cl$_6$ (3\% Bi/Sb incorporation).}
    \label{Fig3}
\end{figure}

Figure \ref{Fig3} shows the photoluminescence excitation and emission spectra under 375 nm (3.30 eV) excitation. All samples exhibit similar broad emission profiles centered around 448 nm. The emission full width at half maximum (FWHM) varies with doping, as tabulated in Table \ref{plt}: pristine Cs$_2$TiCl$_6$ shows FWHM of 153.8 nm, Bi-incorporated systems show 138.2 nm, while Cs$_2$TiCl$_6$ (2\% Sb) exhibits the largest FWHM of 164.73 nm, approximately 20 nm higher than the corresponding Bi-doped sample. 

The broad PL emission indicates self-trapped exciton (STE) formation due to strong exciton-phonon coupling. Photogenerated electron-hole pairs localize within isolated [TiCl$_6$]$^{2-}$ octahedra, forming emissive STE states. Hybrid DFT calculations by Kavanagh et al. \cite{pltheo} confirm strongly bound excitons in Cs$_2$TiX$_6$ systems arising from isolated octahedra.

 The broader FWHM in Sb-doped samples could indicate enhanced electron-phonon coupling, but may also arise from structural inhomogeneity due to dopant-induced local distortions, or multiple emitting sites with slightly different environments.

\begin{table}[h]
\centering   
\caption{FWHM of PL emission for each sample}
\label{plt}
\begin{tabular}{lc}
\hline\hline
Sample & FWHM (nm) \\
\hline
Cs$_2$TiCl$_6$ & 153.8 \\
Cs$_2$TiCl$_6$ (3\% Bi) & 138.2 \\
Cs$_2$TiCl$_6$ (3\% Sb) & 152.8 \\
Cs$_2$TiCl$_6$ (2\% Bi) & 144.4 \\
Cs$_2$TiCl$_6$ (2\% Sb) & 164.73 \\
\hline\hline
\end{tabular}
\end{table}

\begin{figure}[htbp]
\centering
\begin{subfigure}{0.33\textwidth}
    \includegraphics[width=\textwidth]{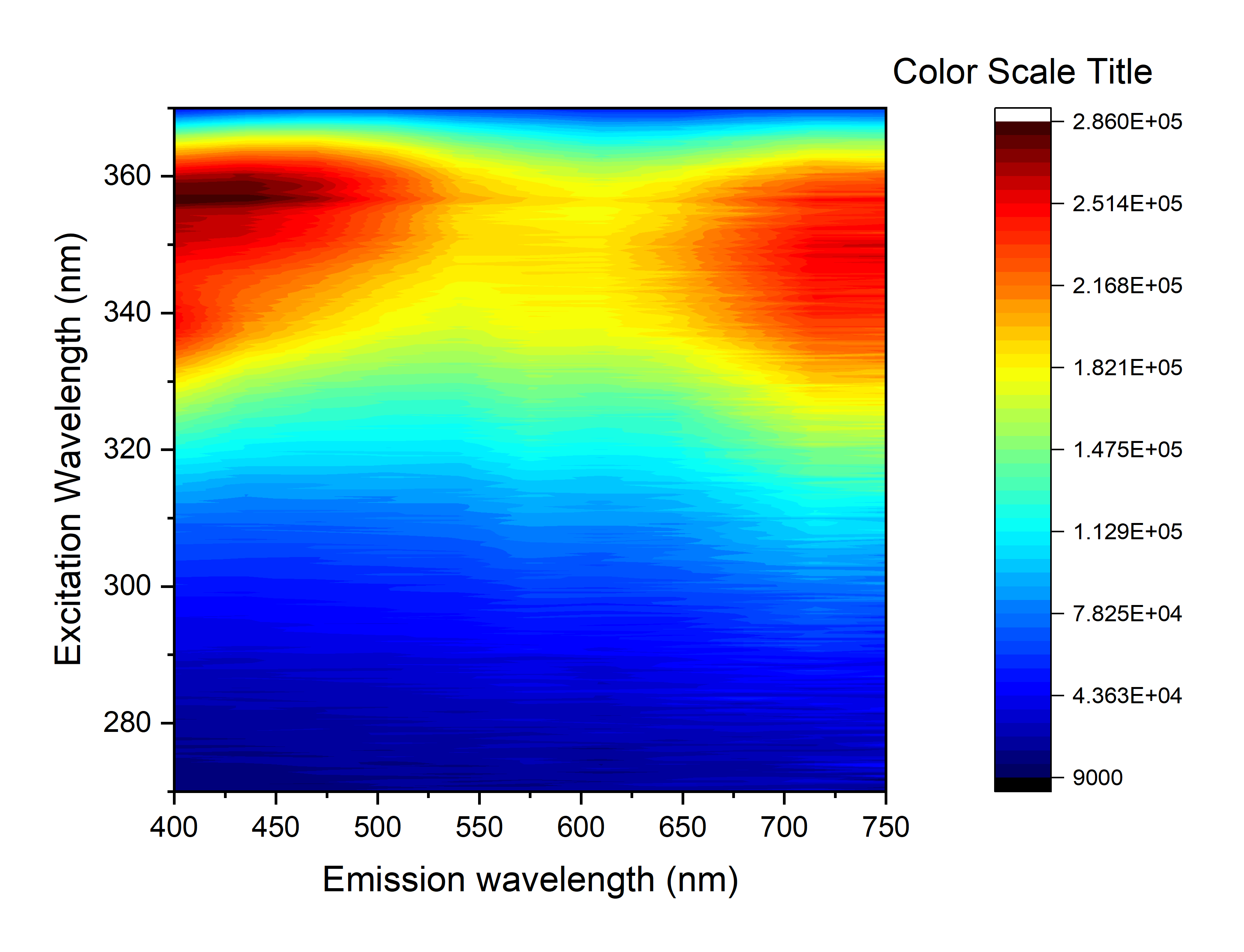}
    \caption{Cs$_2$TiCl$_6$}
\end{subfigure}
\hfill
\begin{subfigure}{0.32\textwidth}
    \includegraphics[width=\textwidth]{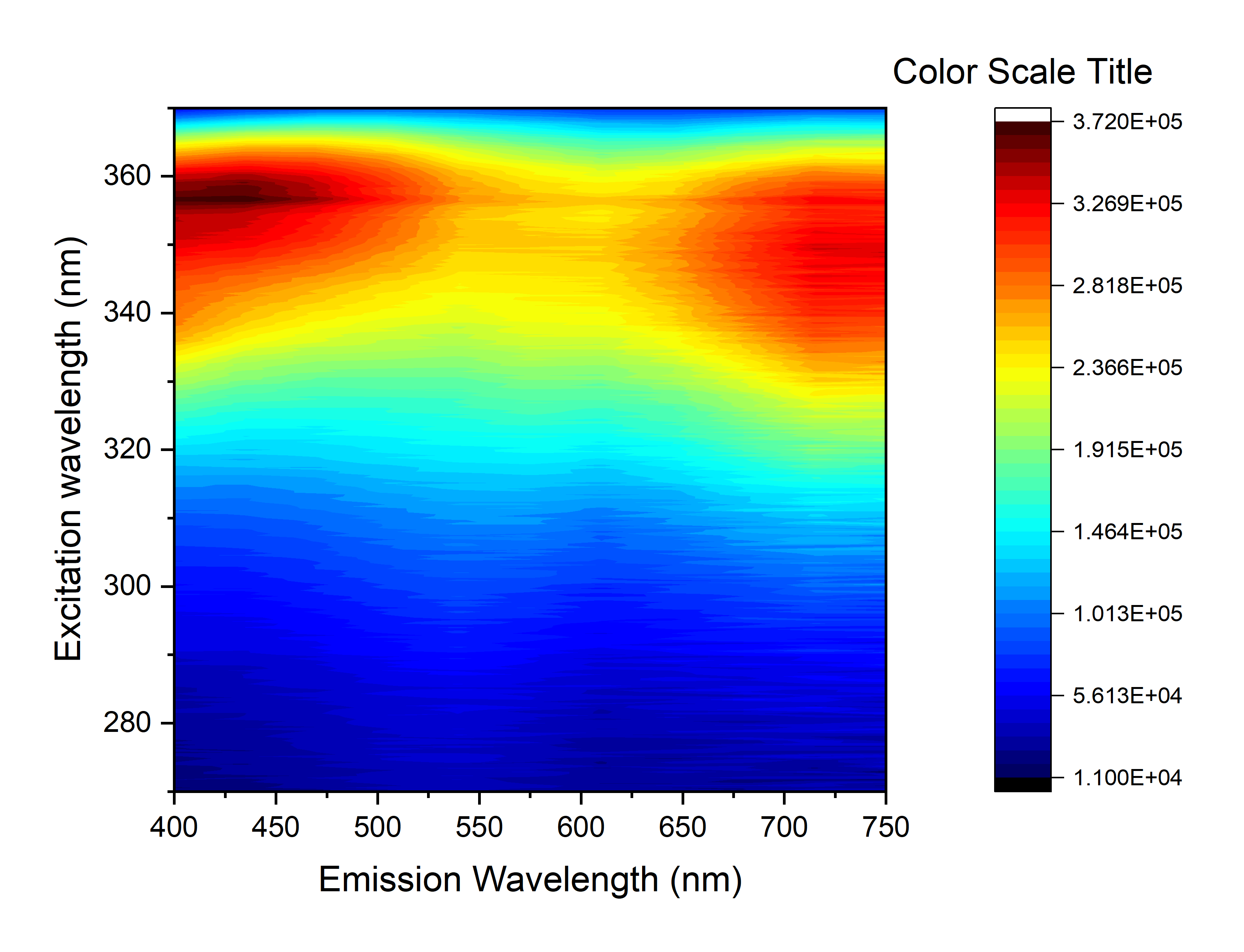}
    \caption{Cs$_2$TiCl$_6$ (3\% Bi)}
\end{subfigure}
\hfill
\begin{subfigure}{0.32\textwidth}
    \includegraphics[width=\textwidth]{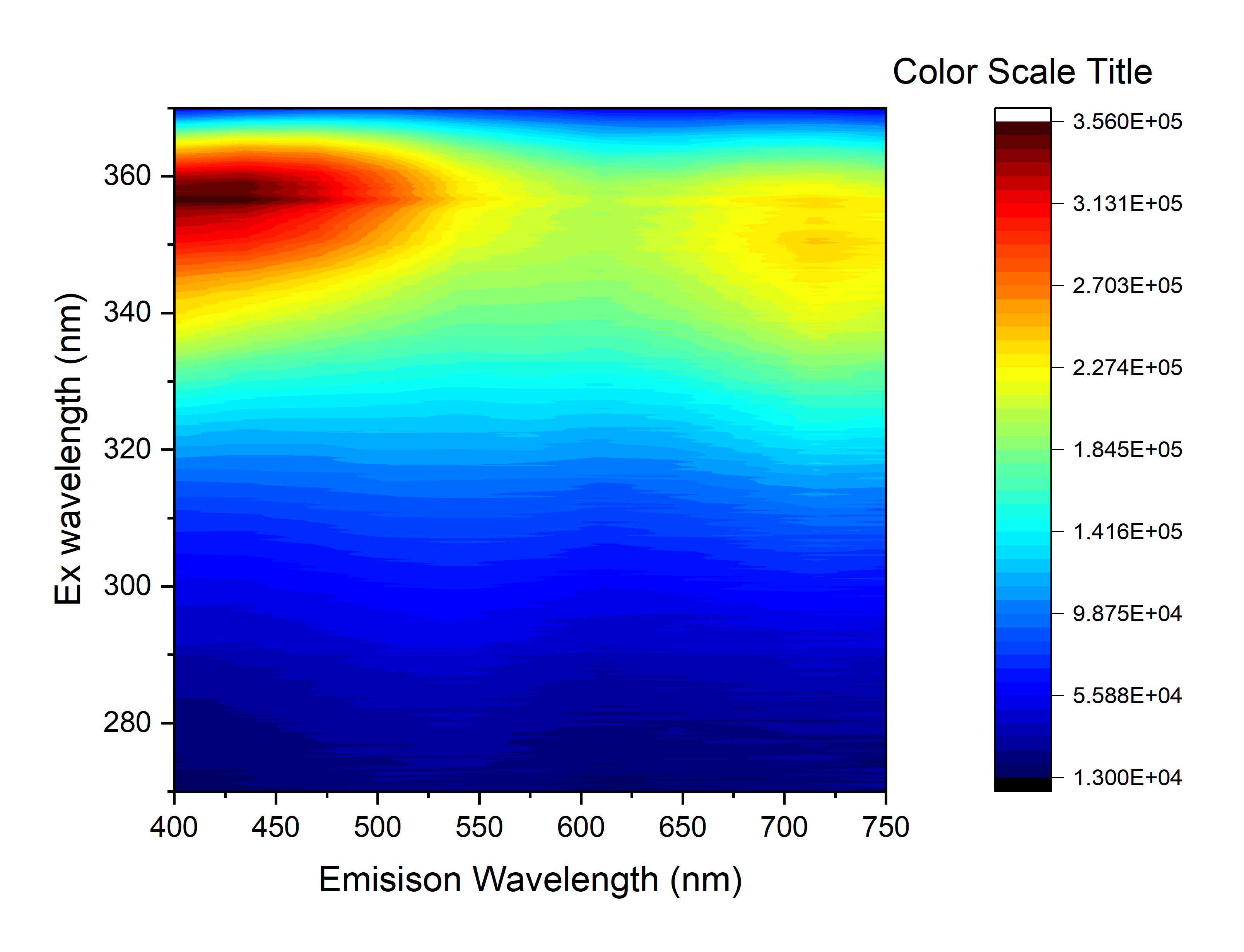}
    \caption{Cs$_2$TiCl$_6$ (3\% Sb)}
\end{subfigure}
     
\caption{Excitation dependent emission of Cs$_2$TiCl$_6$ and Cs$_2$TiCl$_6$ (3\% Sb/Bi incorporated) systems}  
\label{contour}
\end{figure}

The excitation spectrum, recorded with fixed emission at 448 nm, shows a hump at 270-300 nm attributed to Ti-Cl charge transfer transitions within the [TiCl$_6$]$^{2-}$ octahedra. A distinct excitation peak at 315 nm appears in Cs$_2$TiCl$_6$ (2\% Bi) but is absent in Cs$_2$TiCl$_6$ (2\% Sb), corresponding to direct excitation of Bi$^{3+}$ ions from the $^1$S$_0$ ground state to the $^3$P$_1$ excited state. Excitation-dependent emission contour plots (Figure \ref{contour}) reveal maximum PL intensity under 370 nm excitation for all samples. The FWHM increases significantly when shifting to longer excitation wavelengths.

\begin{figure}[htbp]
\centering
 \includegraphics[scale=0.5]{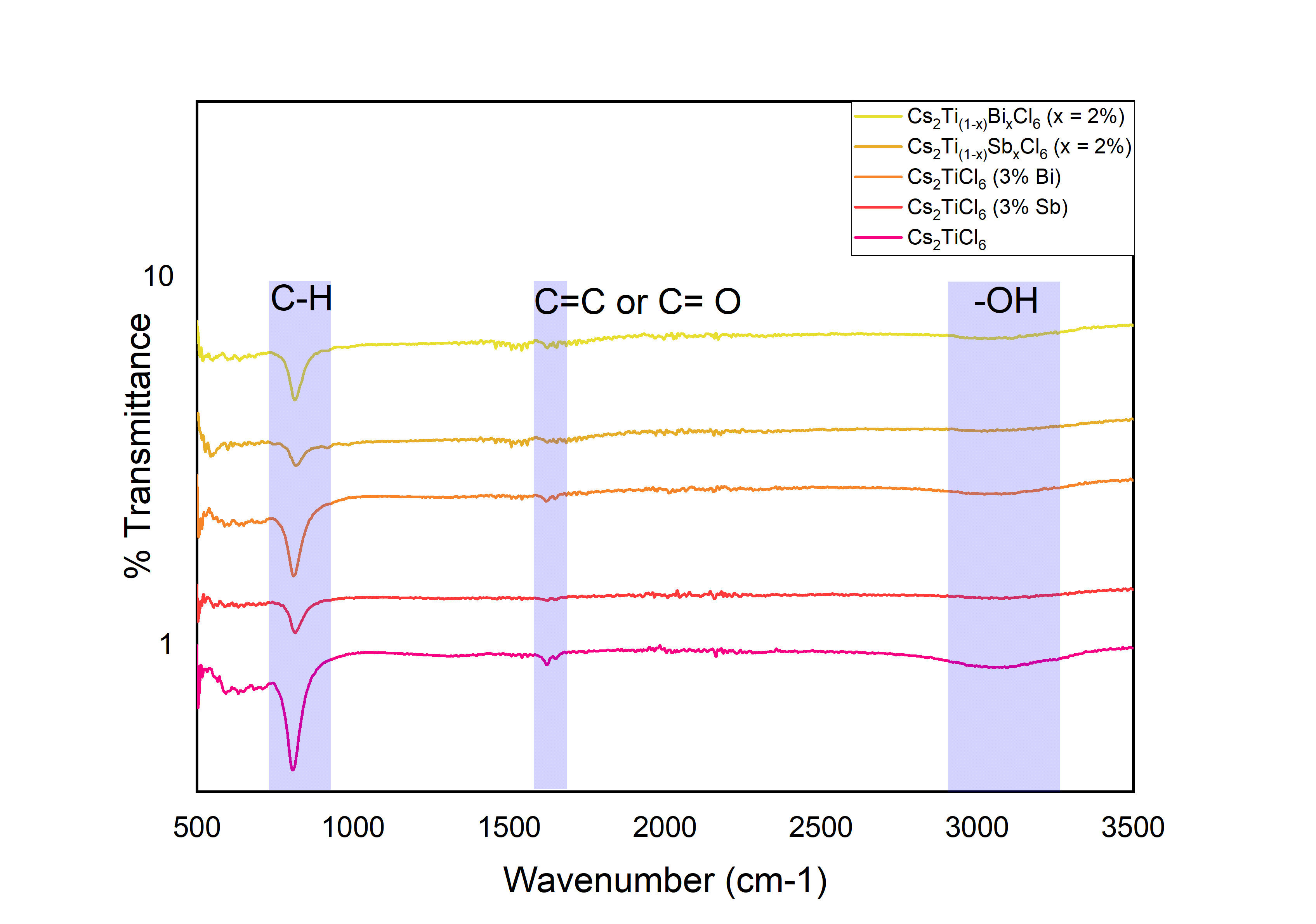}   
 \caption{Ambient FTIR measurement}
 \label{Fig4}
\end{figure}

Fourier-transform infrared spectroscopy (FTIR) as shown in Figure \ref{Fig4} reveals three absorption regions: 750-930 cm$^{-1}$ (C-H bonds), 1590-1640 cm$^{-1}$ (C=C or C=O bonds), and 2910-3260 cm$^{-1}$ (-OH bonds), indicating organic contamination (likely from synthesis solvents) in the samples.

\subsection{Stability Assessment}

\begin{figure}[htbp]
\centering
\begin{subfigure}{0.31\textwidth}
    \includegraphics[width=\textwidth]{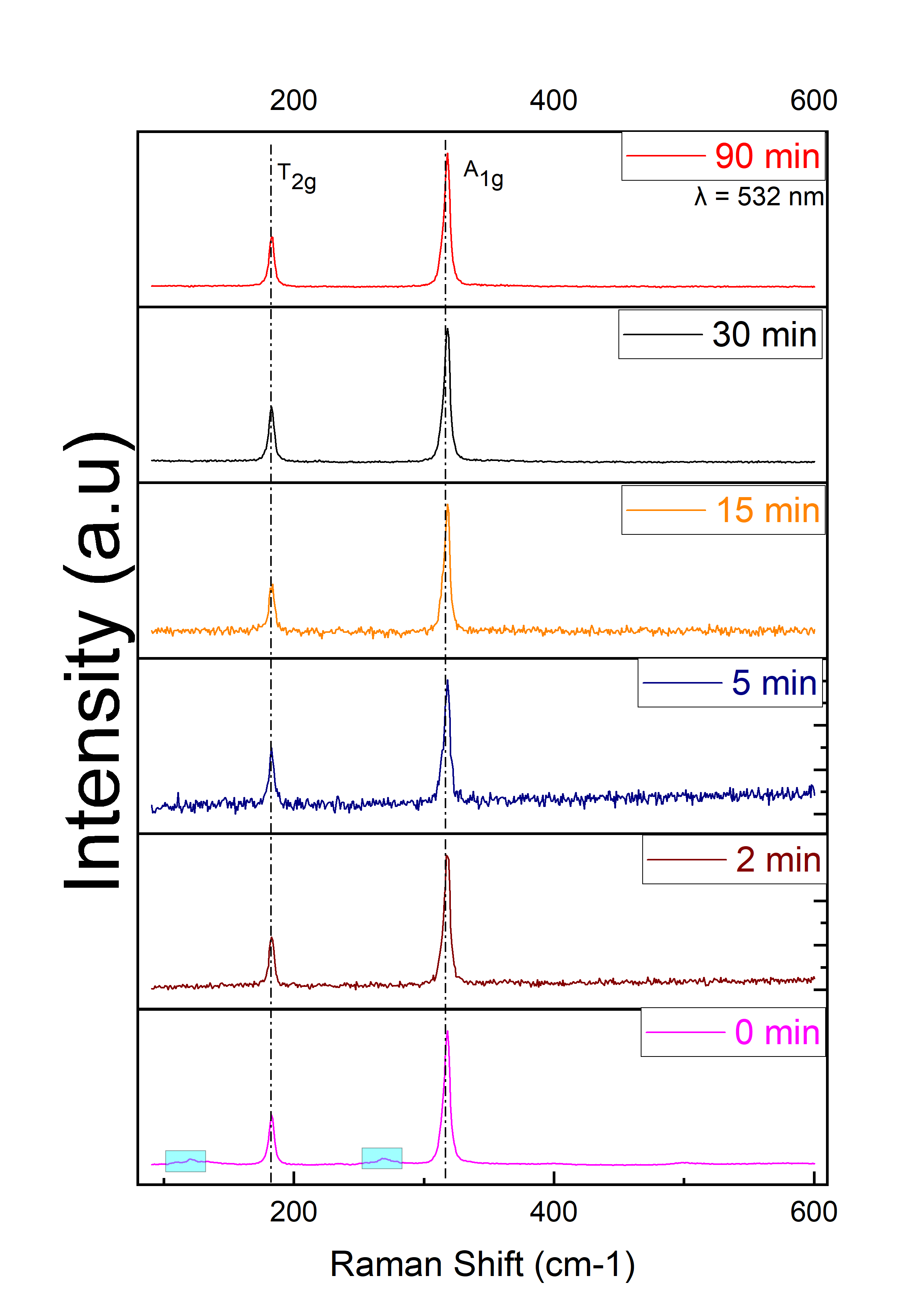}
    \caption{Cs$_2$TiCl$_6$}
\end{subfigure}
\hfill
\begin{subfigure}{0.31\textwidth}
    \includegraphics[width=\textwidth]{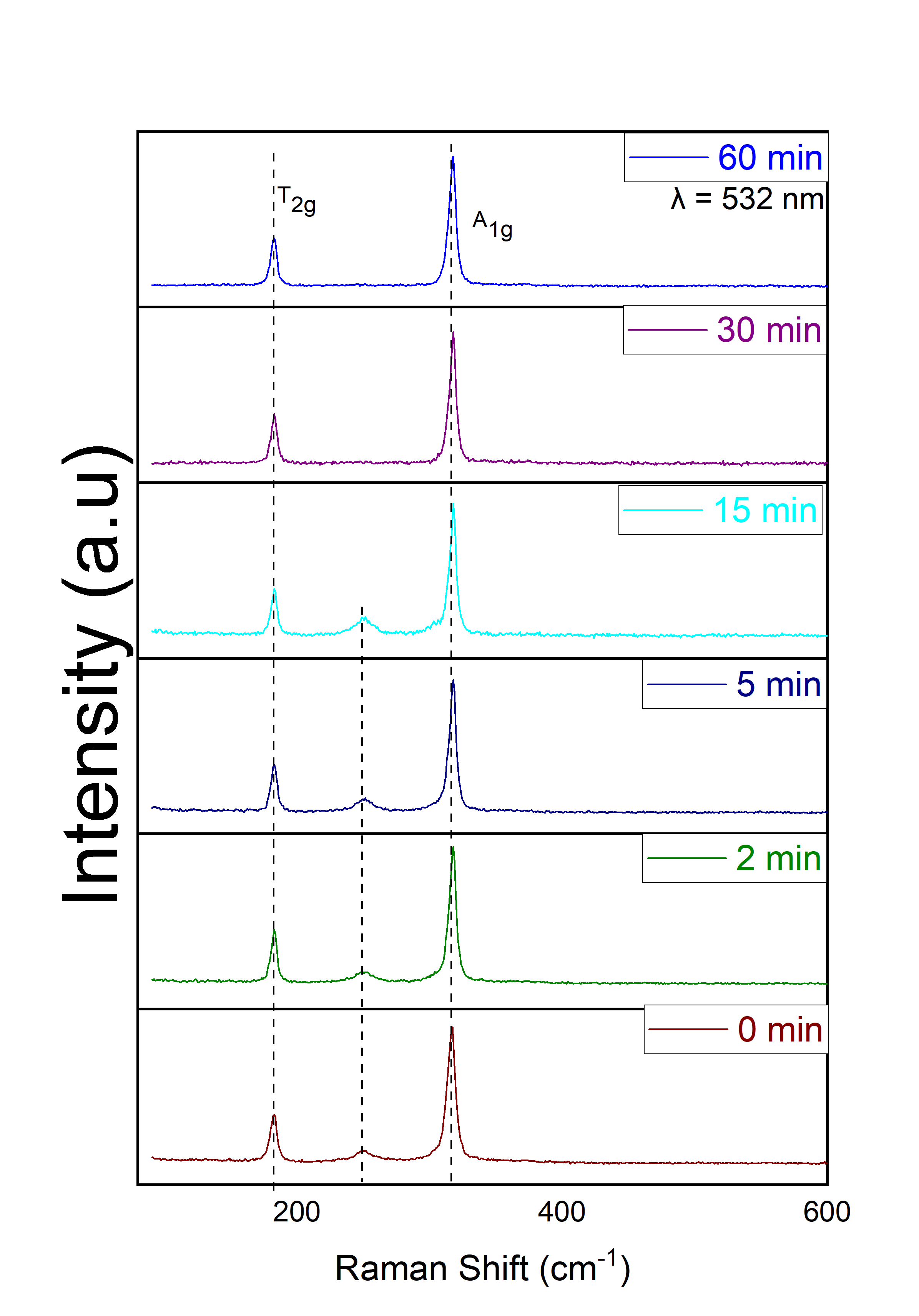}
    \caption{Cs$_2$TiCl$_6$ (3\% Bi)}
\end{subfigure}
\hfill
\begin{subfigure}{0.31\textwidth}
    \includegraphics[width=\textwidth]{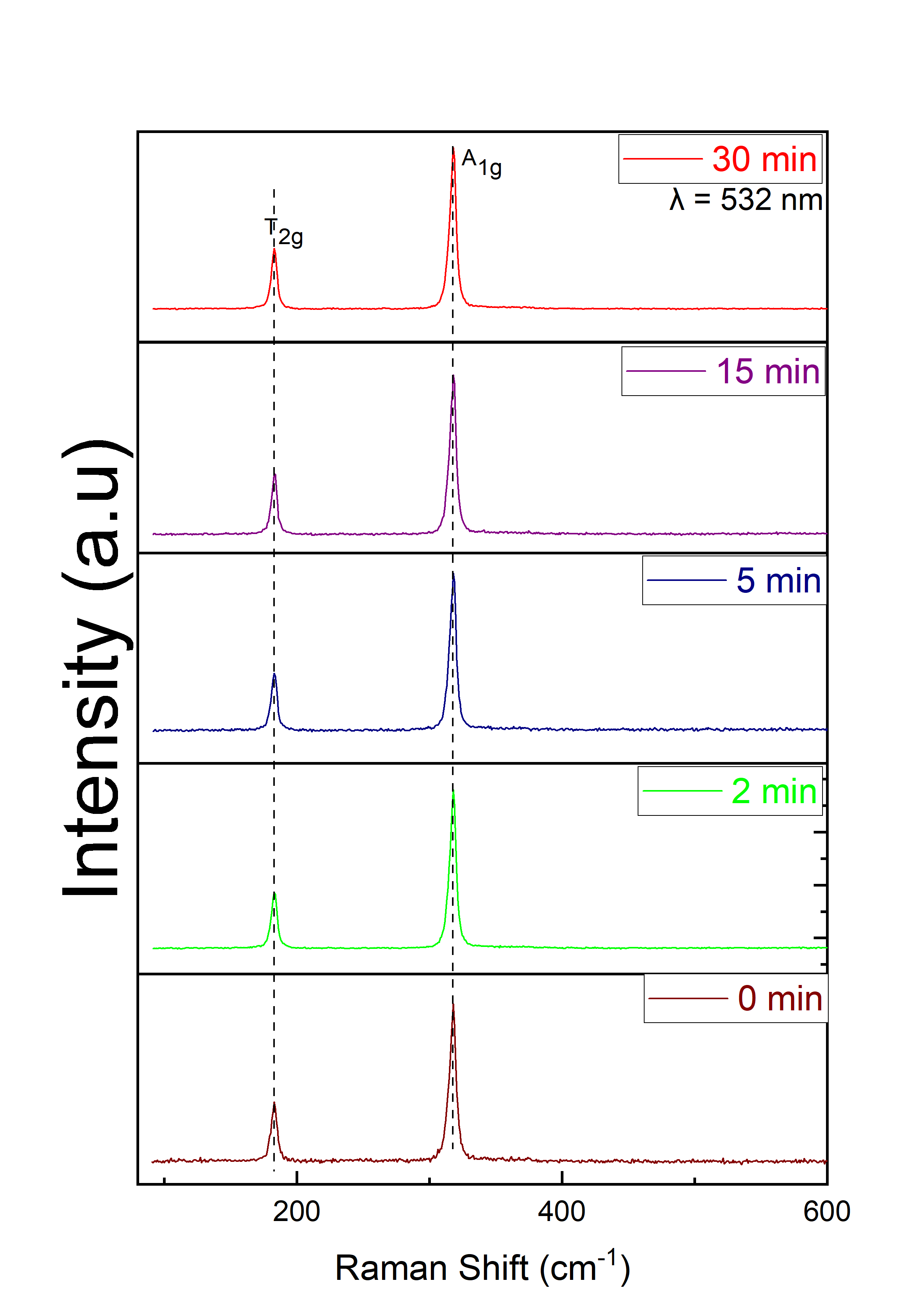}
    \caption{Cs$_2$TiCl$_6$ (3\% Sb)}
\end{subfigure}
       
\caption{Raman spectra evolution of Cs$_2$TiCl$_6$, Cs$_2$TiCl$_6$ (3\% Bi) and Cs$_2$TiCl$_6$ (3\% Sb) systems with time with addition of water}
\label{Fig5}
\end{figure}

\begin{figure}[htbp]
\centering
 \includegraphics[width=0.6\textwidth]{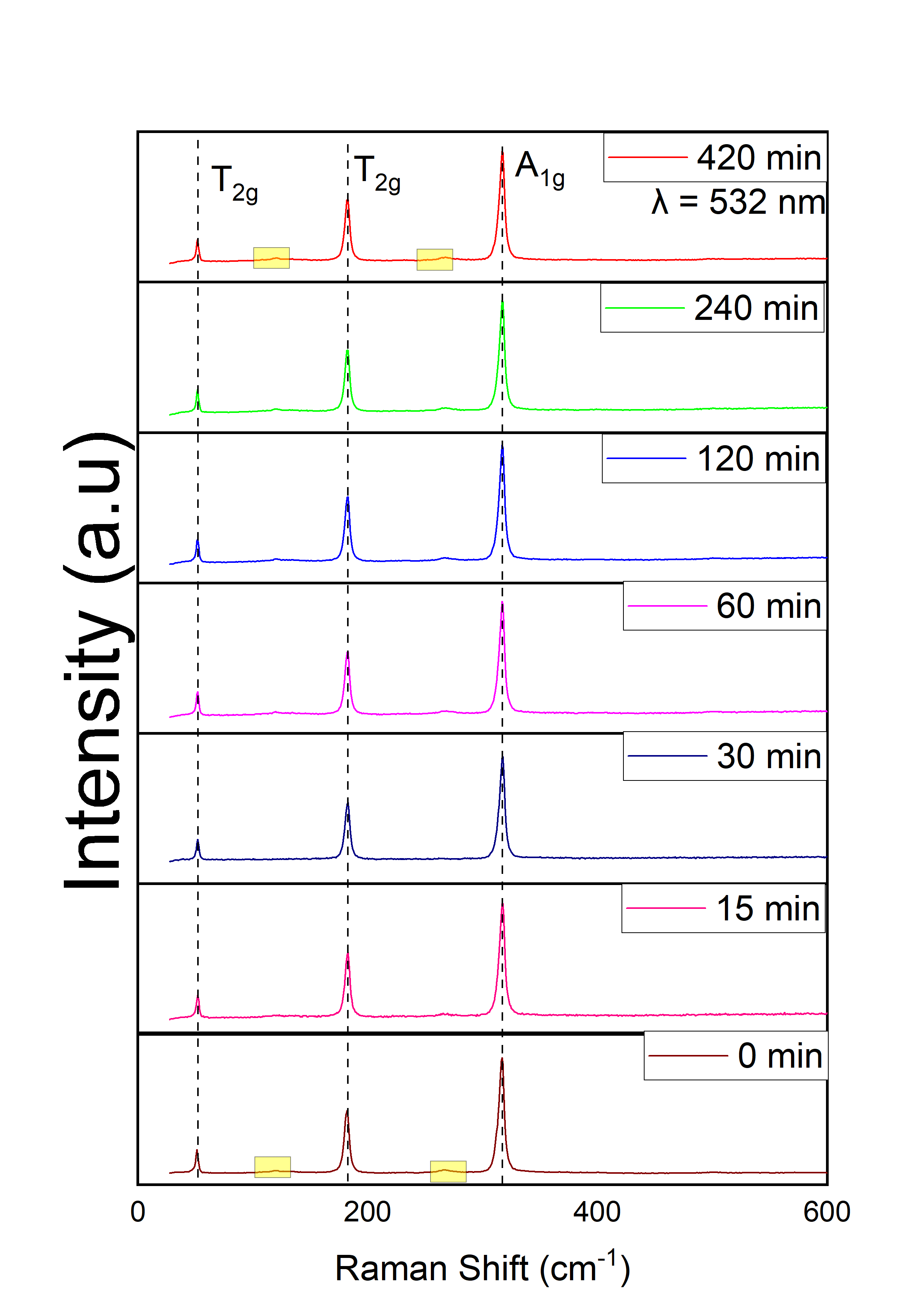}   
 \caption{Evolution of Raman spectra of Cs$_2$TiCl$_6$ on exposure with heat}
 \label{Fig6}
\end{figure}

Water stability was evaluated by adding 1 mL of water to powder samples and monitoring Raman spectra. The Cs$_2$TiCl$_6$ (3\% Sb) system showed no visible changes, with high-intensity T$_{2g}$ and A$_{1g}$ modes at 180 and 318 cm$^{-1}$ remaining intact. However, low-intensity modes at 120 and 268 cm$^{-1}$ disappeared immediately upon water addition in Cs$_2$TiCl$_6$ and after 15 minutes in Cs$_2$TiCl$_6$ (3\% Bi), suggesting these peaks originate from water-sensitive impurity phases, possibly surface-localized rather than bulk impurities. Thermal stability was assessed by heating samples at 120$^{\circ}$C in a vacuum furnace with Raman spectra collected at regular intervals. No spectral changes were observed in Cs$_2$TiCl$_6$ after 7 hours of continuous heating, confirming thermal stability at 120$^{\circ}$C and validating the synthesis drying temperature up to 100$^{\circ}$C. The Raman spectra corresponding to stability assessment are plotted in Figures \ref{Fig5} and \ref{Fig6}.

\subsection{Temperature-Dependent Raman Spectroscopy}

\begin{figure}[htbp]
\centering
 \includegraphics[width=0.7\textwidth]{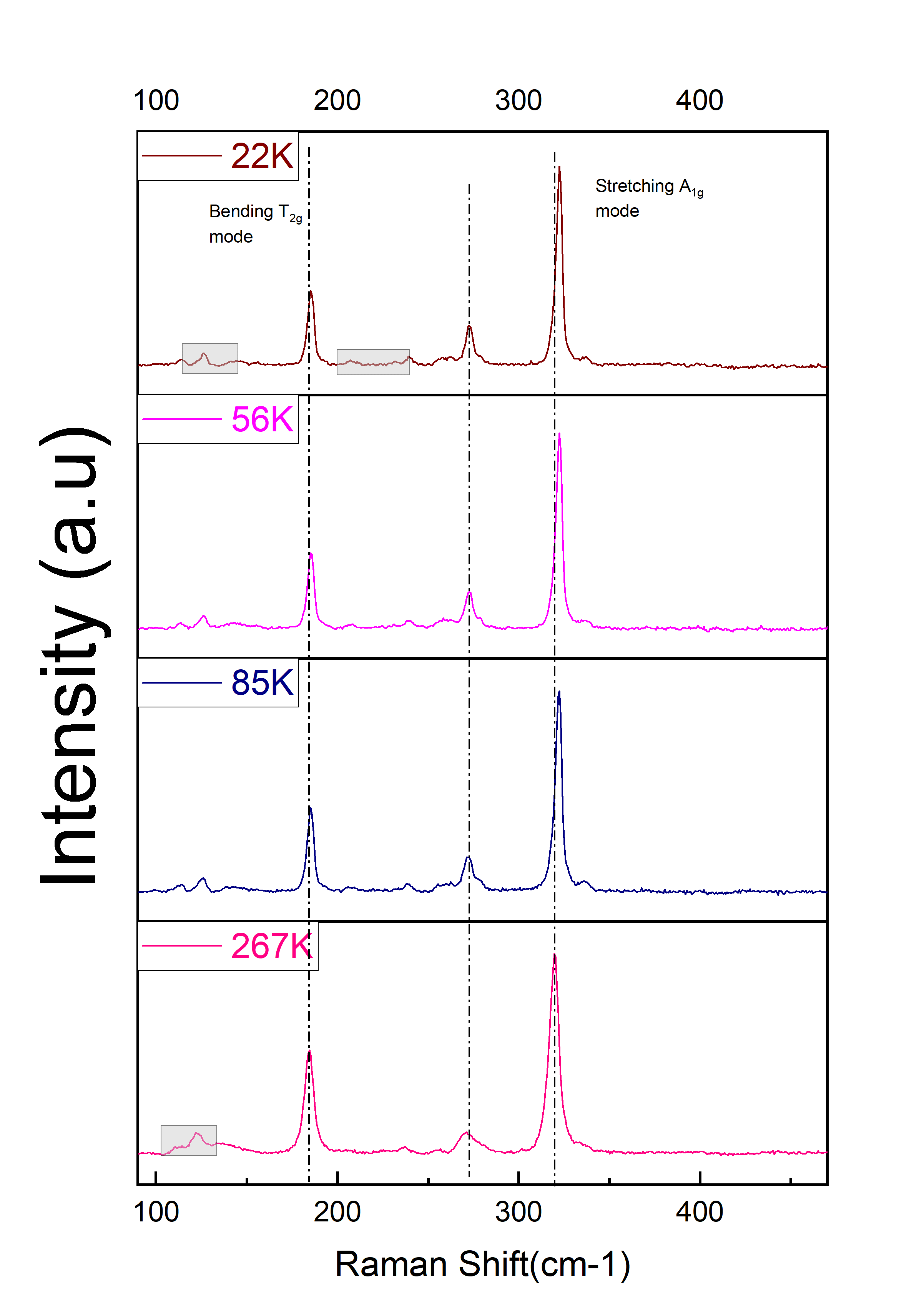}   
 \caption{Temperature dependent Raman spectra evolution of Cs$_2$TiCl$_6$ system}
 \label{Fig7}
\end{figure}

Low-temperature Raman measurements were performed on Cs$_2$TiCl$_6$, Cs$_2$TiCl$_6$ (3\% Sb), and Cs$_2$TiCl$_6$ (2\% Sb) from 4 K to 273 K in the spectral range 20-500 cm$^{-1}$ as depicted in Figures \ref{Fig7}, \ref{Fig9}, and \ref{Fig12}.

For pristine Cs$_2$TiCl$_6$, five Raman modes are observed at room temperature: three identified modes corresponding to Cs$^+$ translational T$_{2g}$ (52 cm$^{-1}$), bending T$_{2g}$ (182 cm$^{-1}$), and stretching A$_{1g}$ (318 cm$^{-1}$) vibrations within [TiCl$_6$]$^{2-}$ octahedra, and unidentified peaks at 120, 256, and 270 cm$^{-1}$ likely arising from impurity phases. Upon cooling, all peaks sharpen significantly, indicating increased phonon lifetimes, and shift to higher frequencies. Both the 3\% and 2\% Sb-doped samples exhibit only three clean Raman modes at room temperature, consistent with reduced impurity content observed in XRD. 

\textbf{Key observation:} Both Sb-containing samples exhibit a new mode (M$_1$) appearing below 100 K near the A$_{1g}$ stretching mode at $\sim$314-319 cm$^{-1}$. This mode is absent in pristine Cs$_2$TiCl$_6$ and appears to be specific to Sb-doped samples.

\begin{figure}[htbp]
\centering
\begin{subfigure}{0.48\textwidth}
    \includegraphics[width=\textwidth]{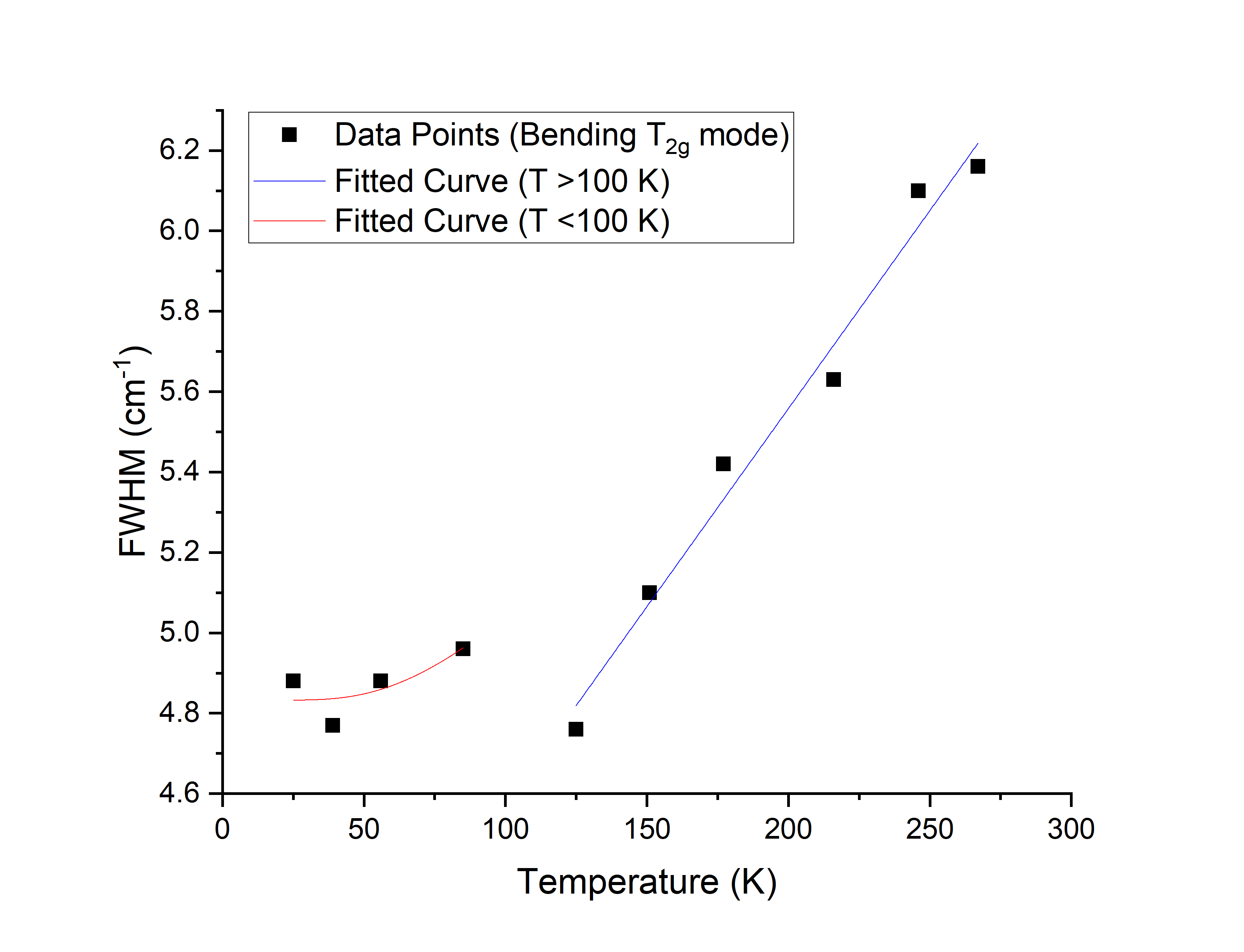}
    \caption{T$_{2g}$ mode: FWHM vs Temperature}
\end{subfigure}
\hfill
\begin{subfigure}{0.48\textwidth}
    \includegraphics[width=\textwidth]{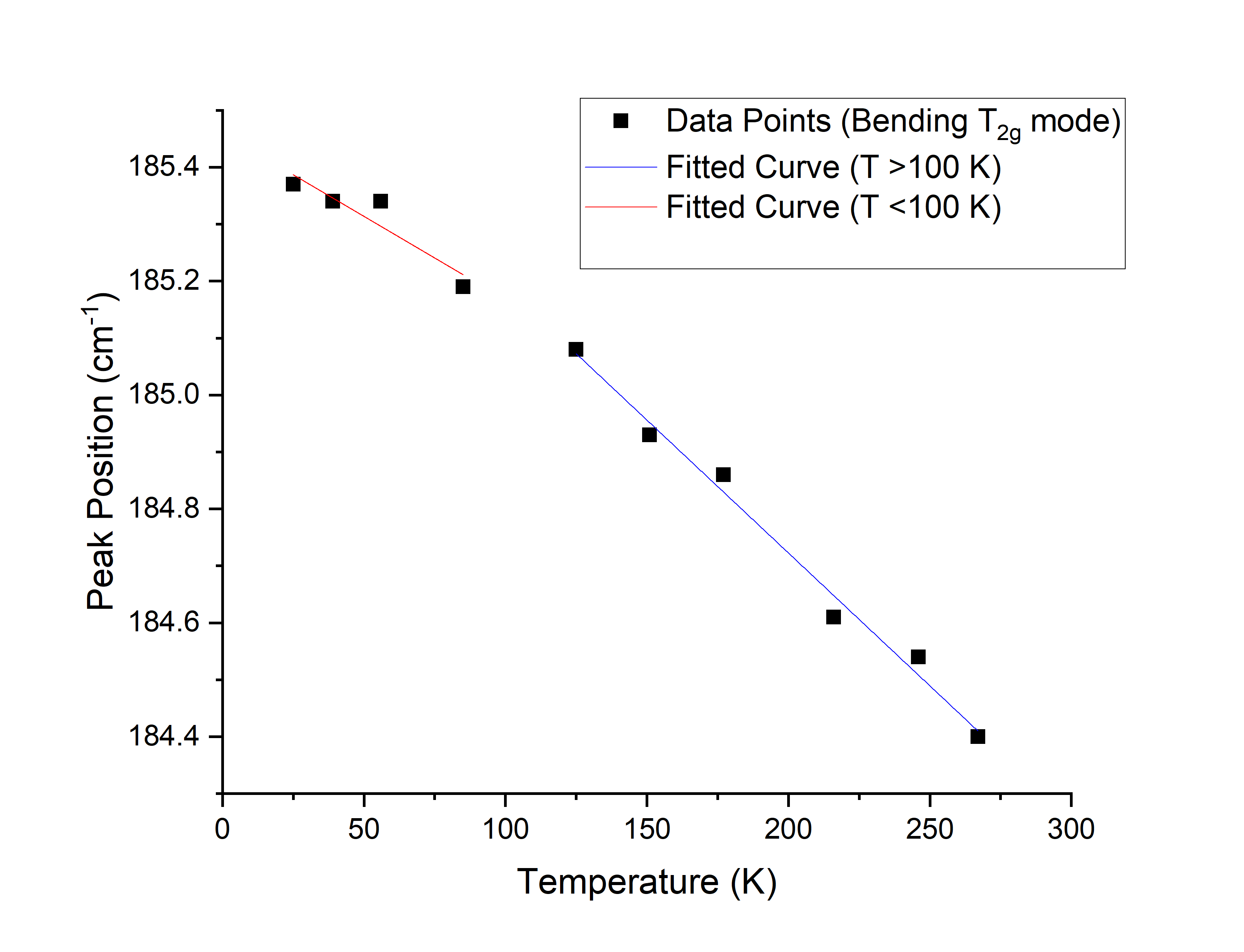}
    \caption{T$_{2g}$ mode: Peak position vs Temperature}
\end{subfigure}

\vspace{0.5cm}

\begin{subfigure}{0.48\textwidth}
    \includegraphics[width=\textwidth]{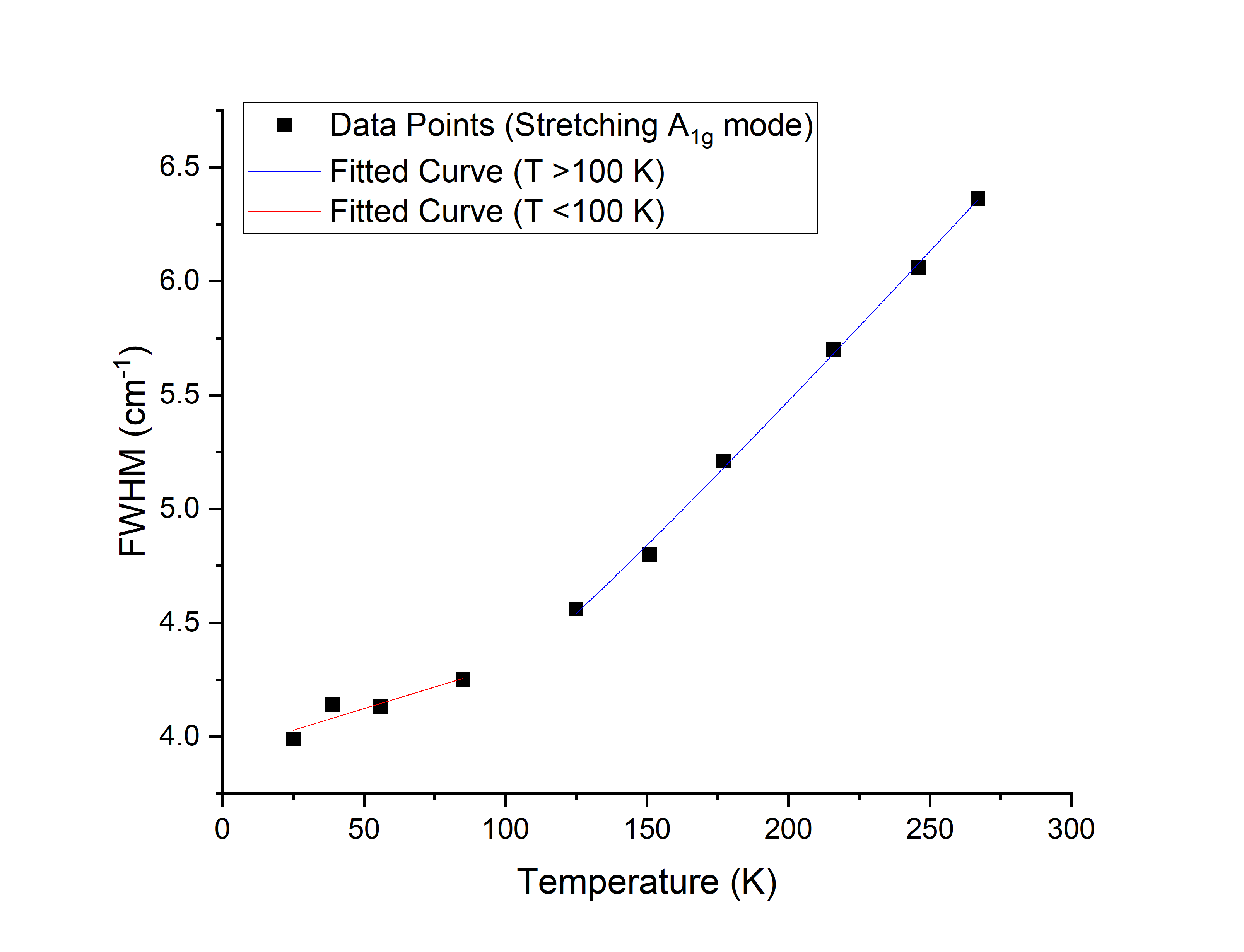}
    \caption{A$_{1g}$ mode: FWHM vs Temperature}
\end{subfigure}
\hfill
\begin{subfigure}{0.48\textwidth}
    \includegraphics[width=\textwidth]{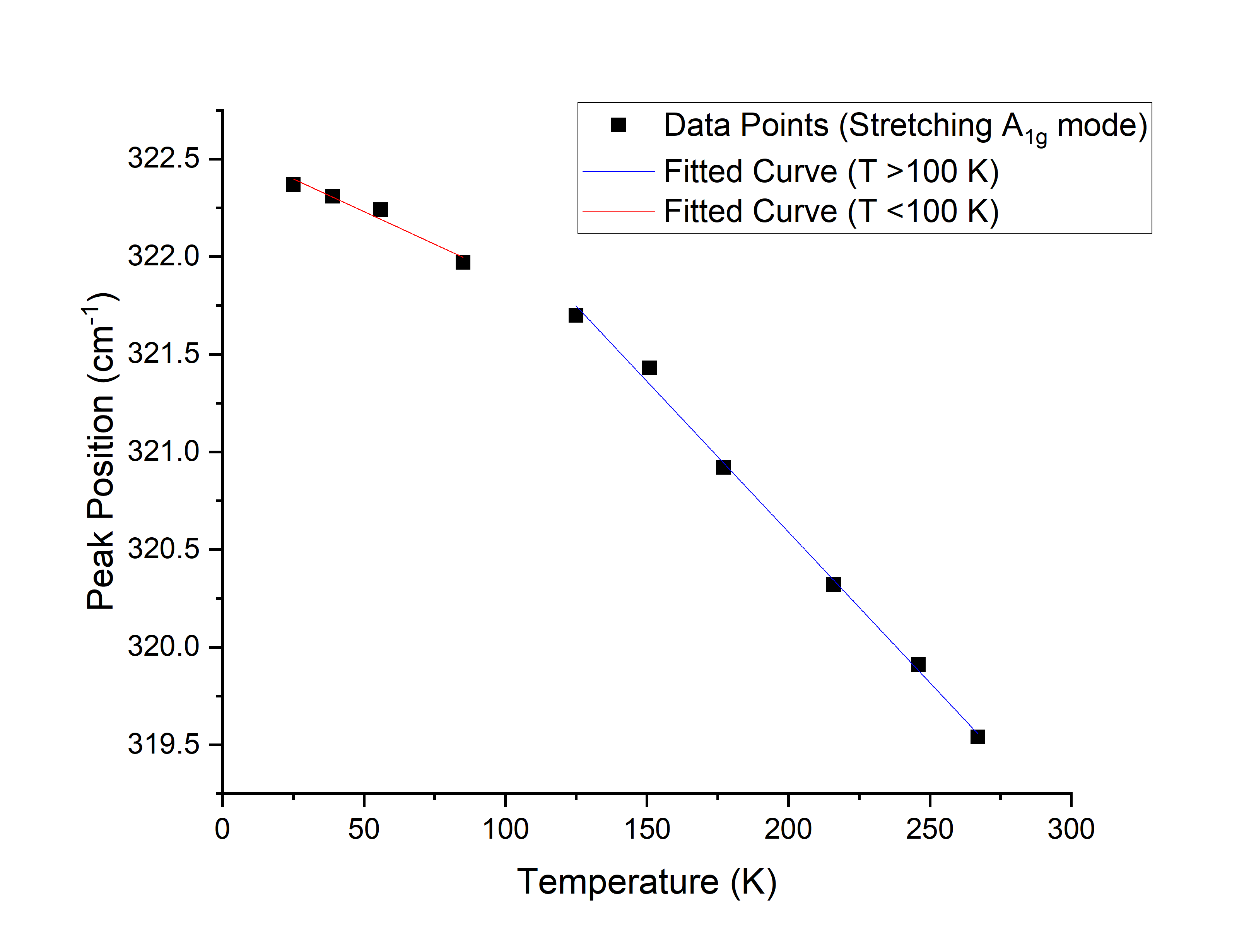}
    \caption{A$_{1g}$ mode: Peak position vs Temperature}
\end{subfigure}
     
\caption{Temperature dependence of FWHM and peak position for bending T$_{2g}$ and stretching A$_{1g}$ modes of Cs$_2$TiCl$_6$. Solid lines represent fits to Eqs. (1) and (2).}  
\label{Fig8}
\end{figure}

\begin{figure}[htbp]
\centering
 \includegraphics[width=0.6\textwidth]{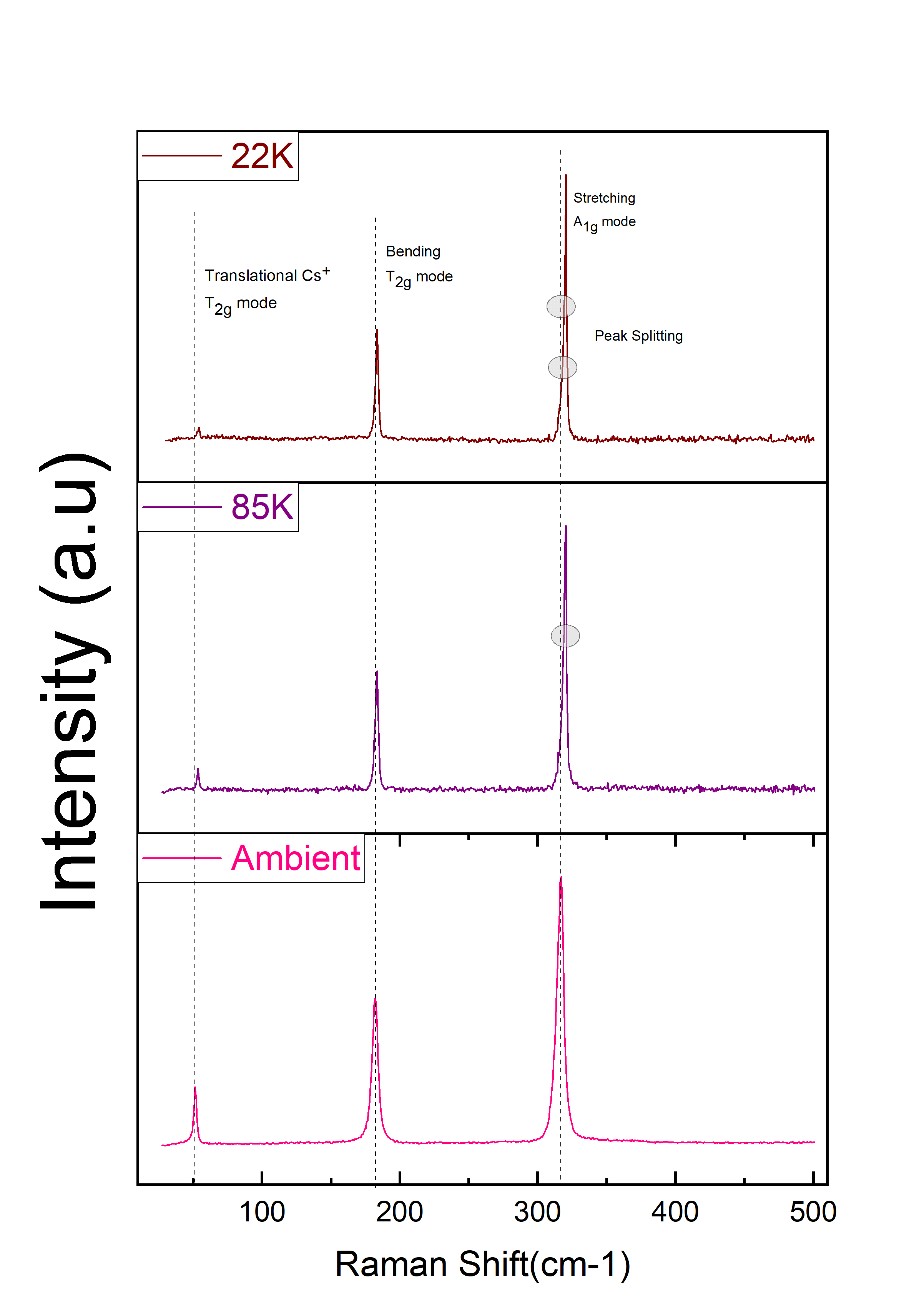}   
 \caption{Raman spectra of Cs$_2$TiCl$_6$ (3\% Sb incorporated) system at Ambient, 85K and 22K}
 \label{Fig9}
\end{figure}

\begin{figure}[htbp]
\centering
\begin{subfigure}{0.48\textwidth}
    \includegraphics[width=\textwidth]{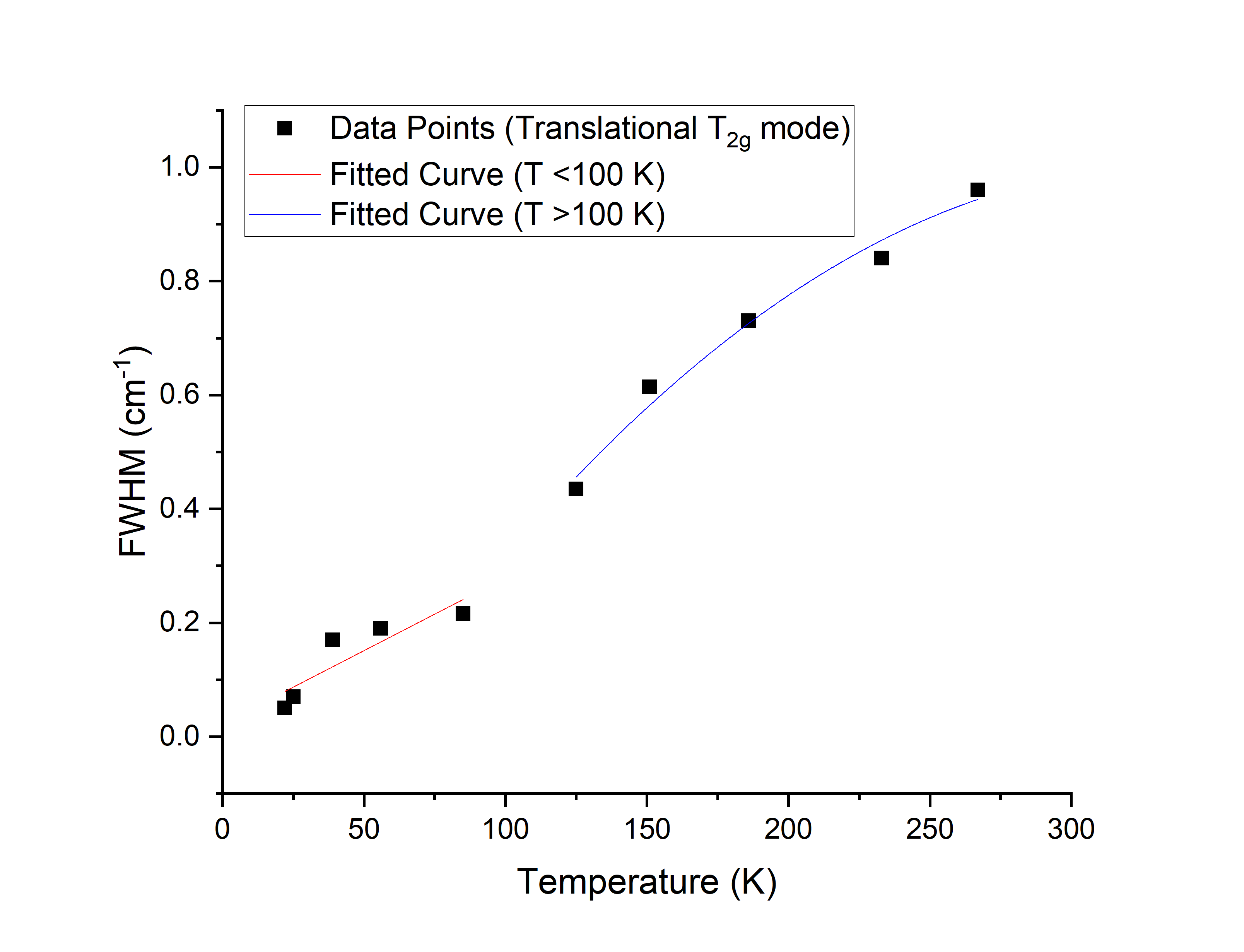}
    \caption{Translational T$_{2g}$: FWHM}
\end{subfigure}
\hfill
\begin{subfigure}{0.48\textwidth}
    \includegraphics[width=\textwidth]{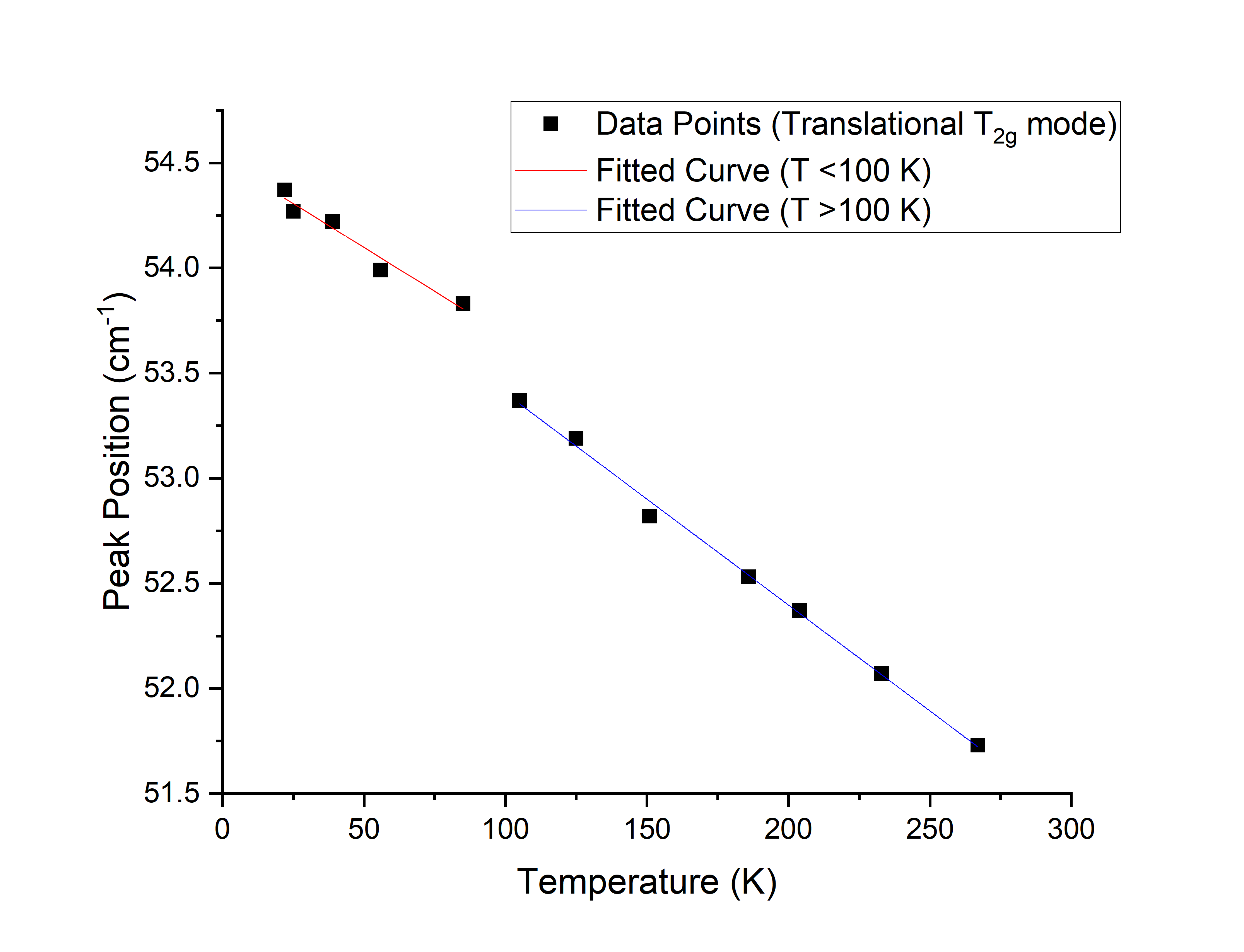}
    \caption{Translational T$_{2g}$: Position}
\end{subfigure}

\vspace{0.3cm}

\begin{subfigure}{0.48\textwidth}
    \includegraphics[width=\textwidth]{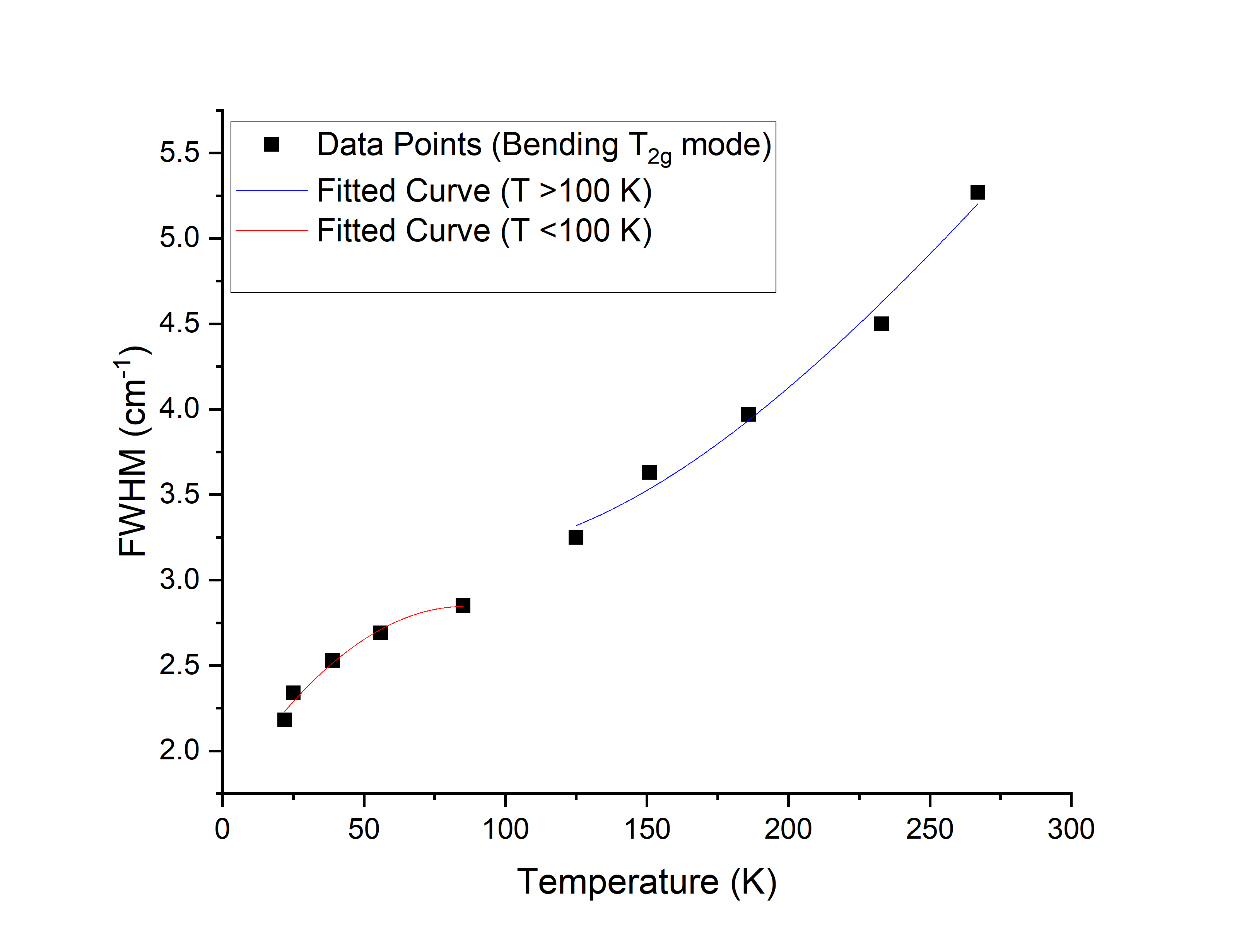}
    \caption{Bending T$_{2g}$: FWHM}
\end{subfigure}
\hfill
\begin{subfigure}{0.48\textwidth}
    \includegraphics[width=\textwidth]{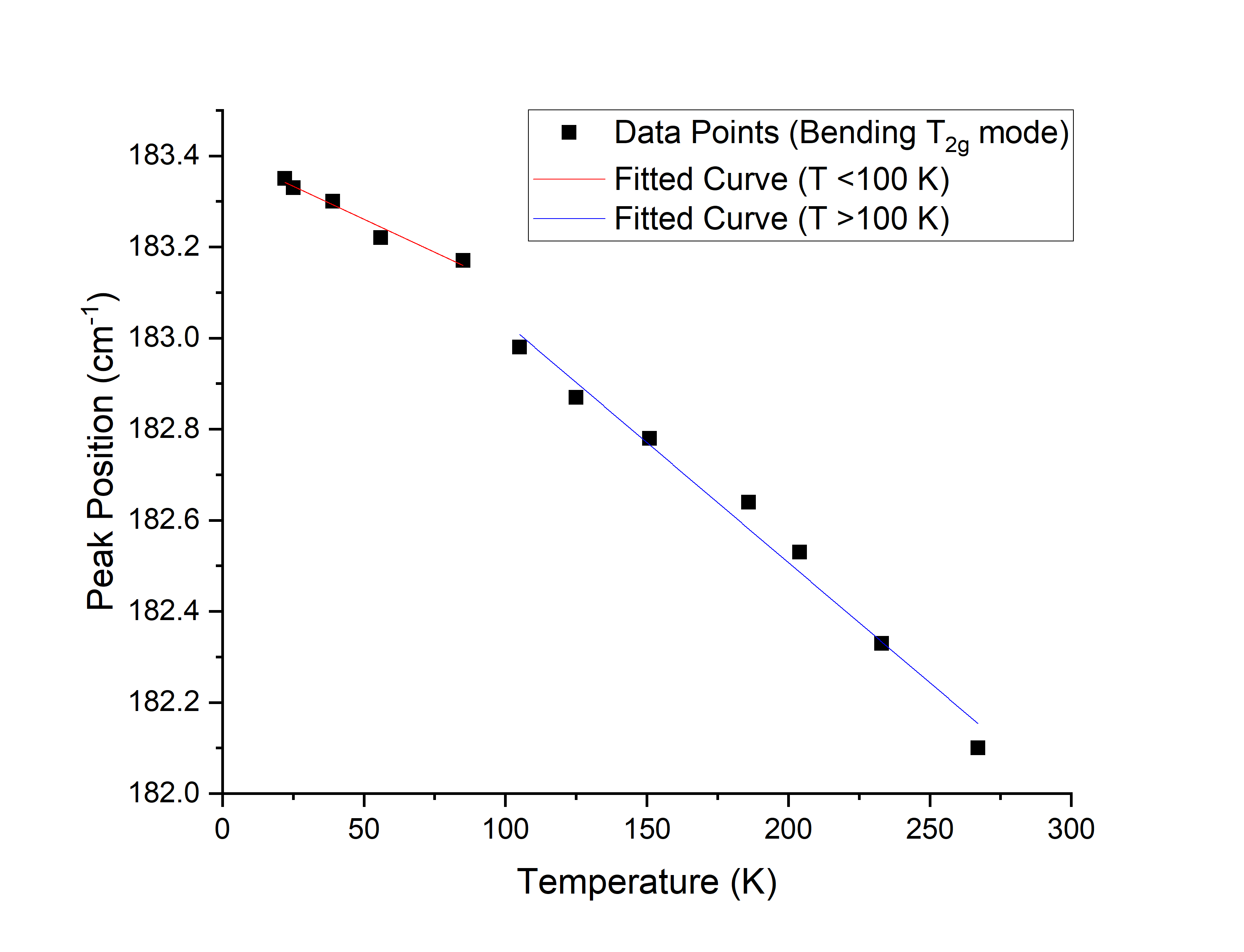}
    \caption{Bending T$_{2g}$: Position}
\end{subfigure}
     
\caption{Temperature dependence of FWHM and peak position for translational and bending T$_{2g}$ modes of Cs$_2$TiCl$_6$ (3\% Sb). Solid lines represent fits to Eqs. (1) and (2).}  
\label{Fig11a}
\end{figure}

\begin{figure}[htbp]
\centering
\begin{subfigure}{0.48\textwidth}
    \includegraphics[width=\textwidth]{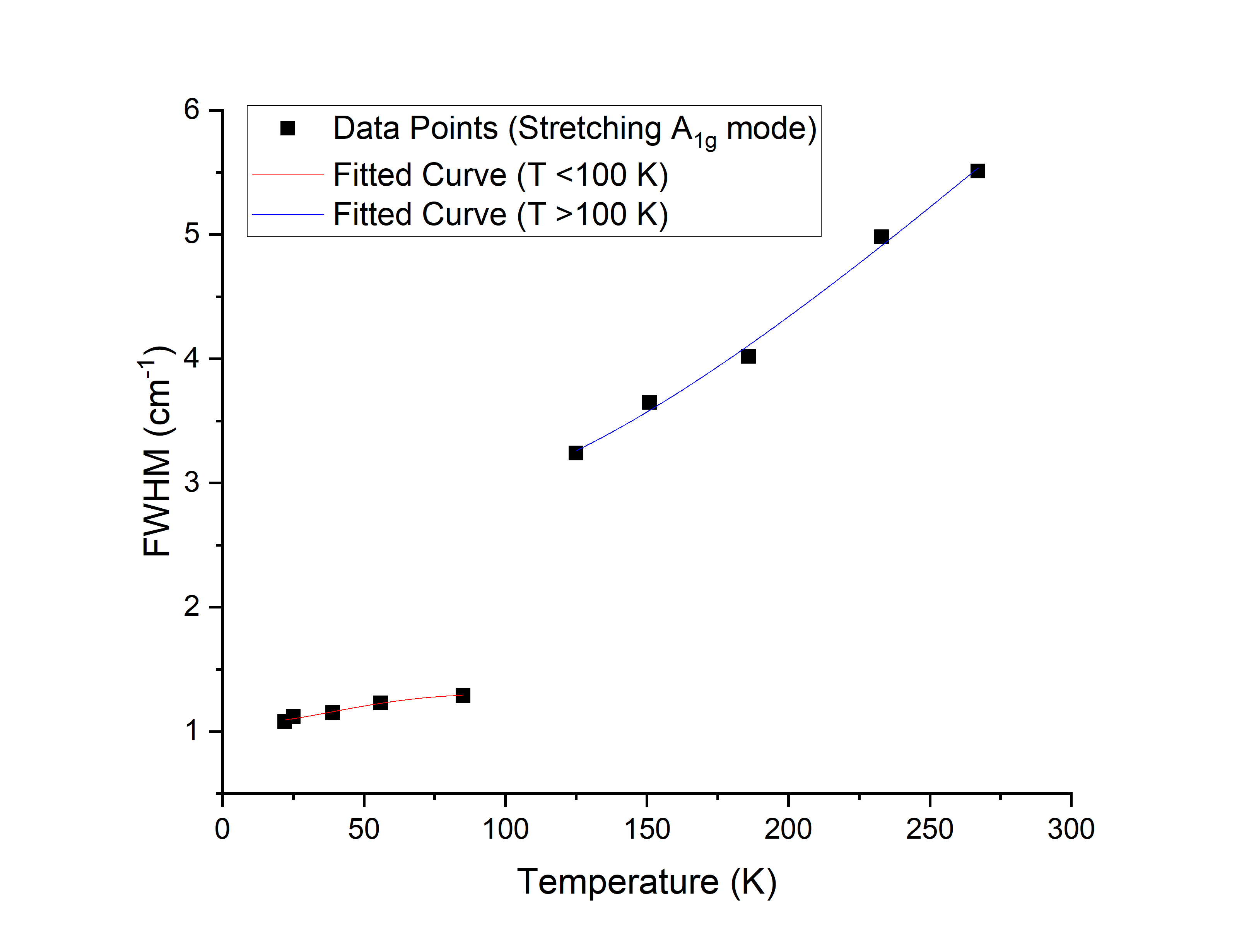}
    \caption{Stretching A$_{1g}$: FWHM}
\end{subfigure}
\hfill
\begin{subfigure}{0.48\textwidth}
    \includegraphics[width=\textwidth]{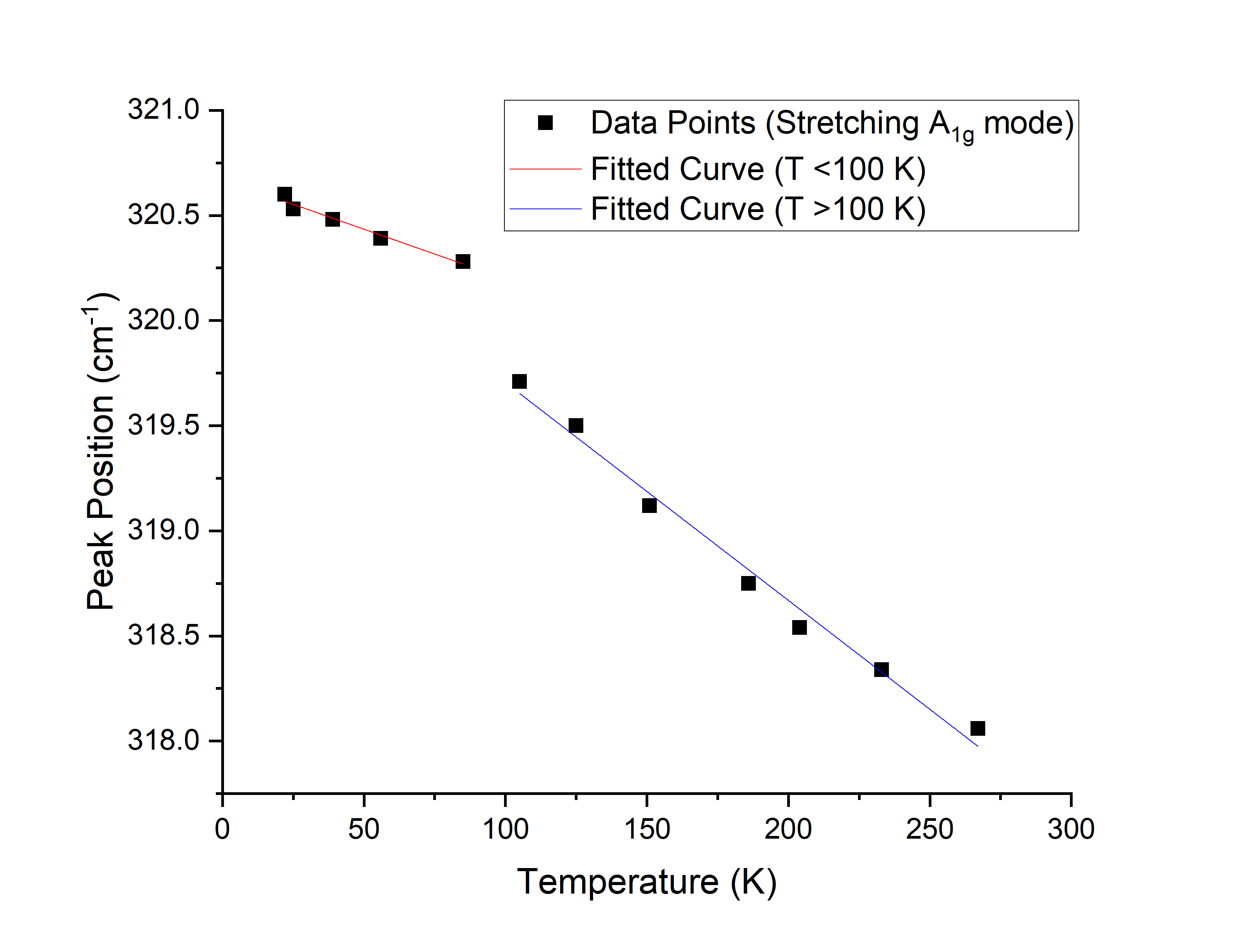}
    \caption{Stretching A$_{1g}$: Position}
\end{subfigure}

\vspace{0.3cm}

\begin{subfigure}{0.48\textwidth}
    \includegraphics[width=\textwidth]{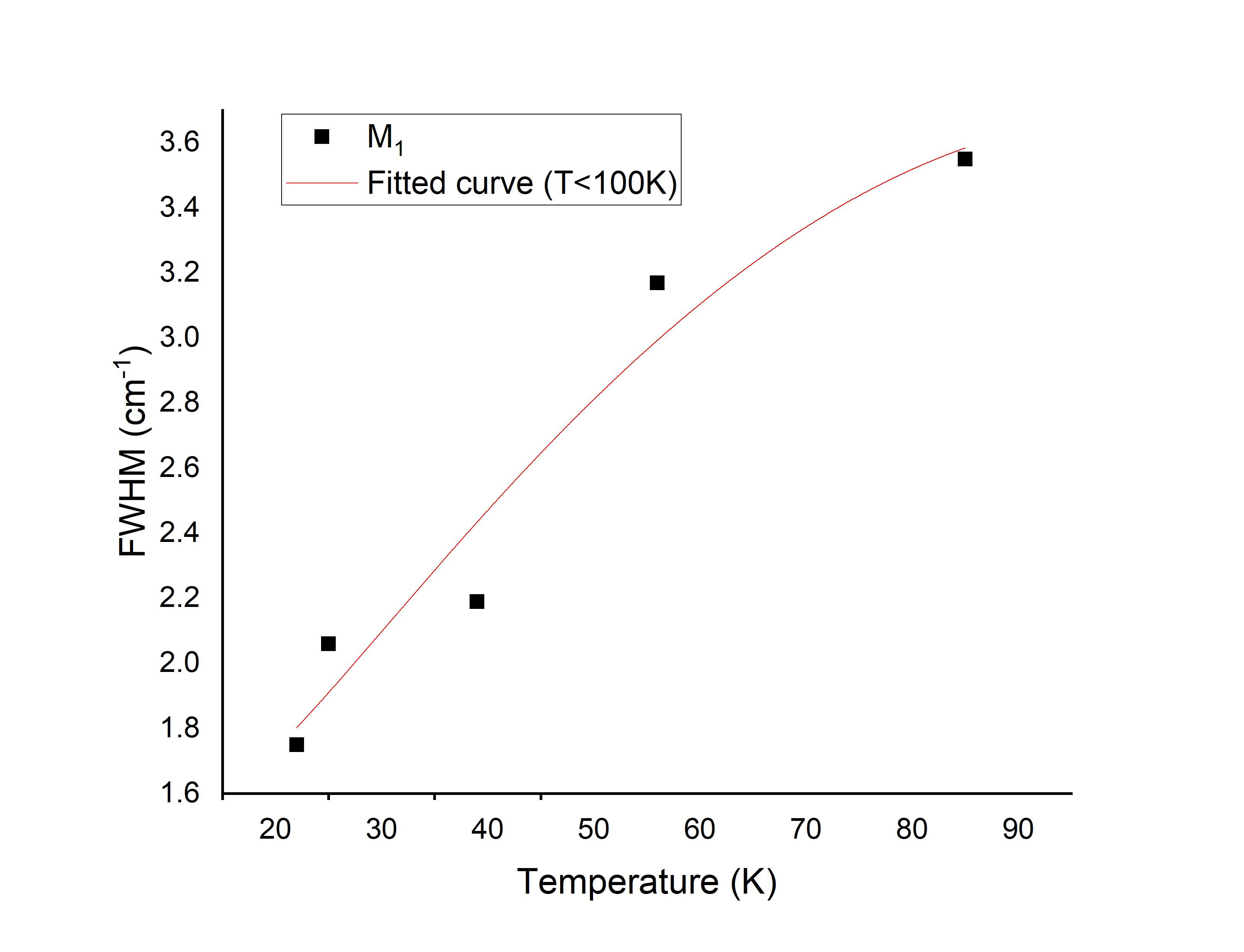}
    \caption{M$_1$ mode: FWHM}
\end{subfigure}
\hfill
\begin{subfigure}{0.48\textwidth}
    \includegraphics[width=\textwidth]{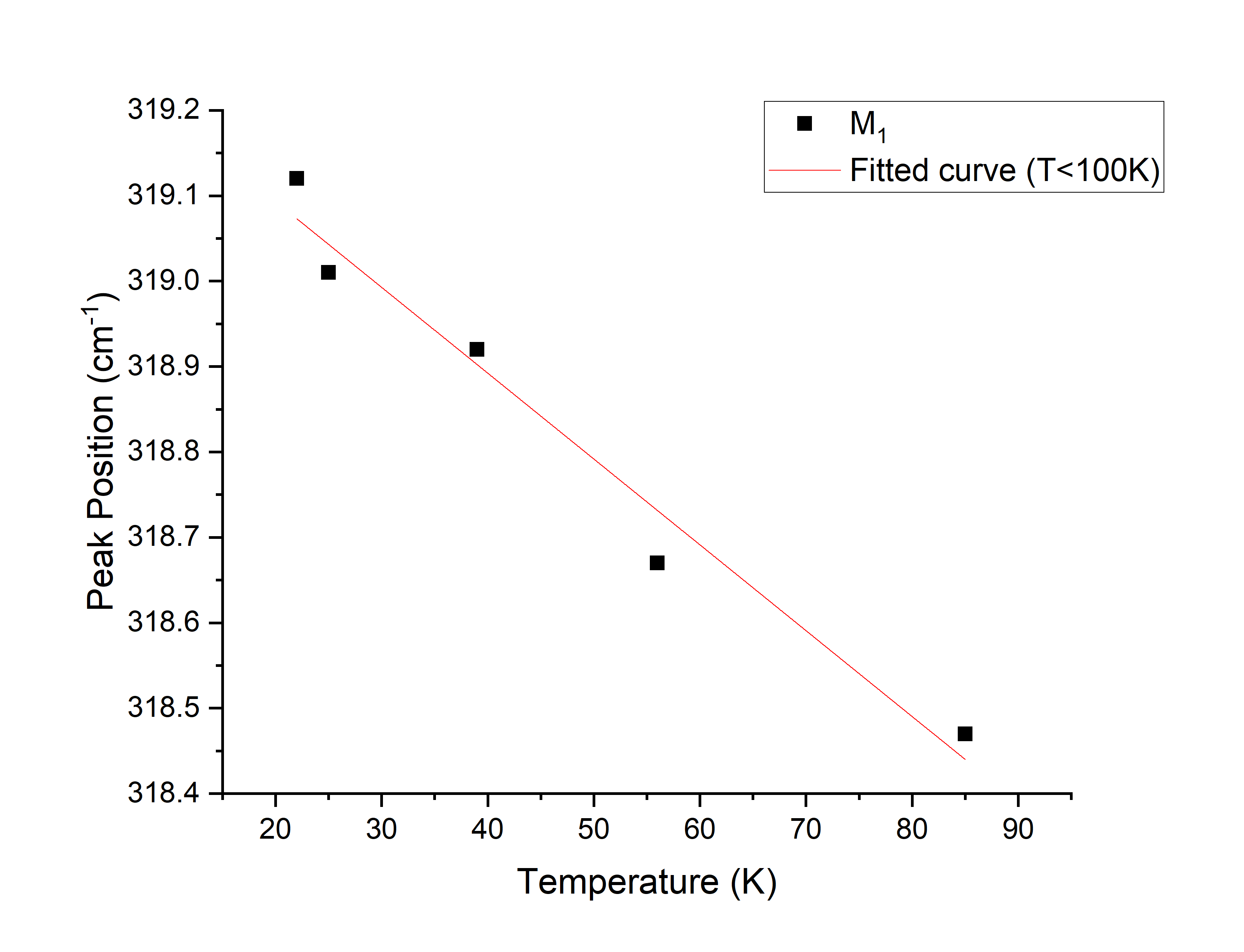}
    \caption{M$_1$ mode: Position}
\end{subfigure}
     
\caption{Temperature dependence of FWHM and peak position for stretching A$_{1g}$ and M$_1$ modes of Cs$_2$TiCl$_6$ (3\% Sb). The M$_1$ mode appears only below 100 K. Solid lines represent fits to Eqs. (1) and (2).}  
\label{Fig11b}
\end{figure}

\begin{figure}[htbp]
\centering
\includegraphics[width=0.7\textwidth]{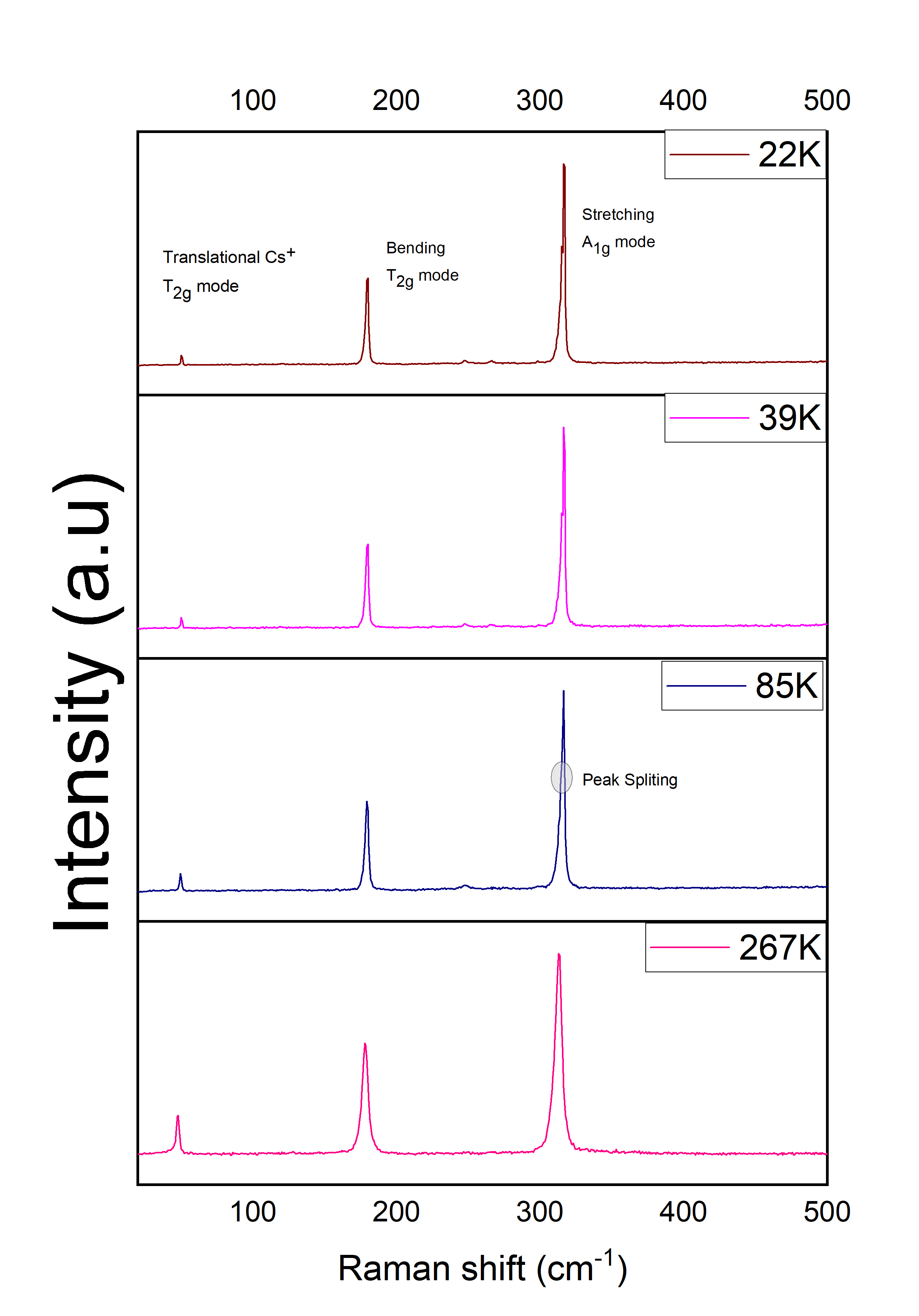}
\caption{Temperature evolution of Raman spectra for Cs$_2$Ti$_{(1-x)}$Sb$_x$Cl$_6$ (x = 2\%) from 4 K to 273 K showing mode hardening and the emergence of M$_1$ mode below 100 K.}  
\label{Fig12}
\end{figure}

\begin{figure}[htbp]
\centering
\begin{subfigure}{0.48\textwidth}
    \includegraphics[width=\textwidth]{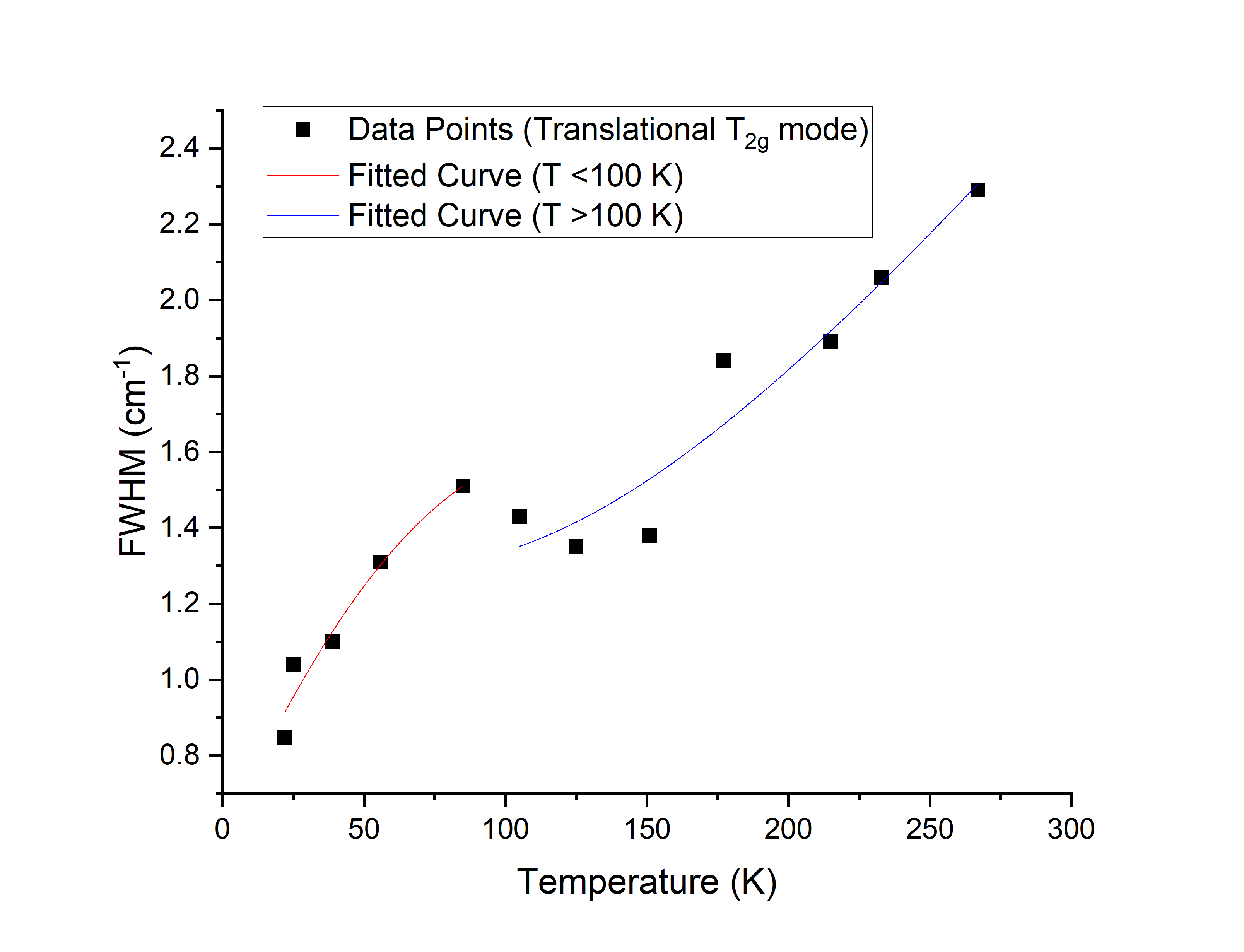}
    \caption{Translational T$_{2g}$: FWHM}
\end{subfigure}
\hfill
\begin{subfigure}{0.48\textwidth}
    \includegraphics[width=\textwidth]{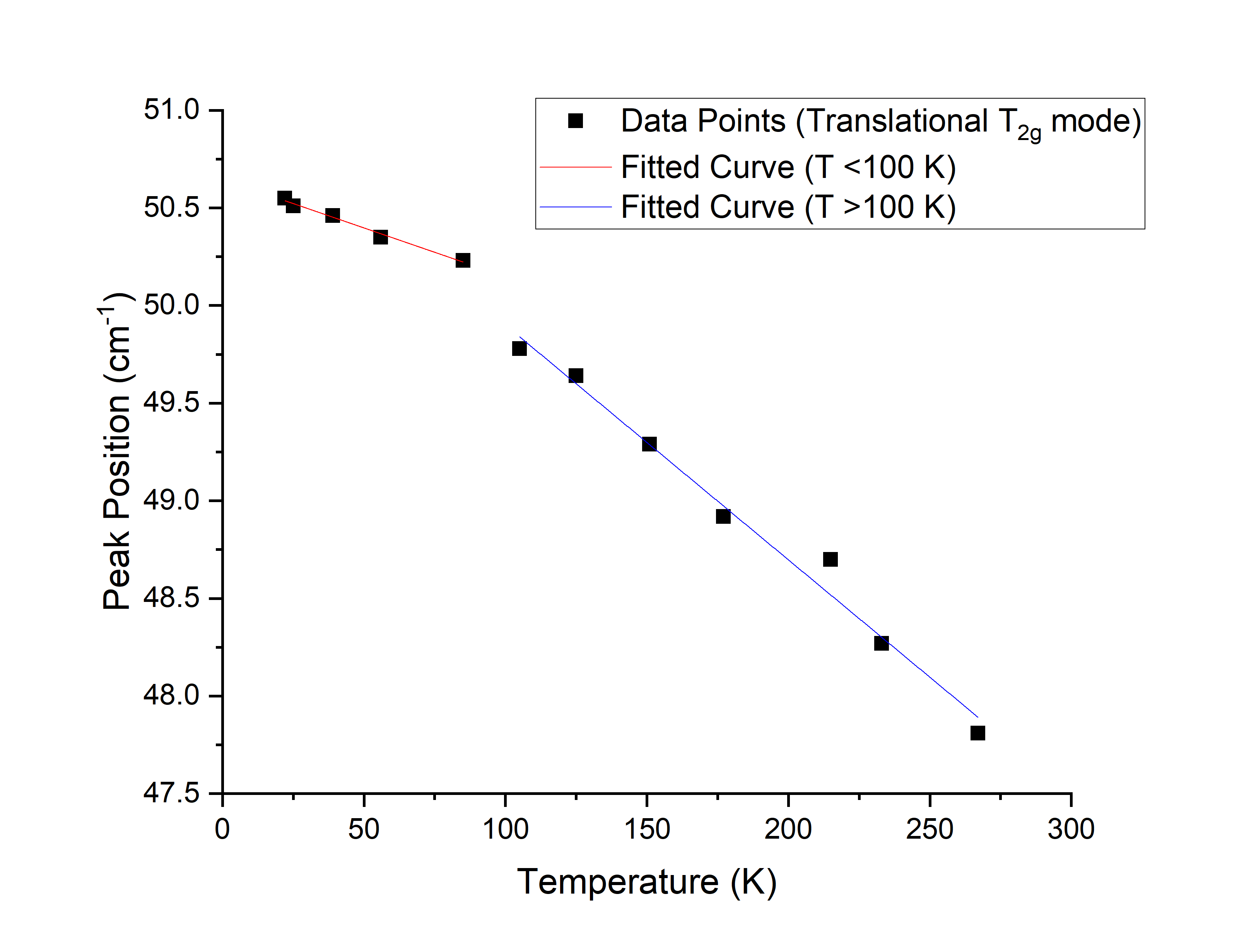}
    \caption{Translational T$_{2g}$: Position}
\end{subfigure}

\vspace{0.3cm}

\begin{subfigure}{0.48\textwidth}
    \includegraphics[width=\textwidth]{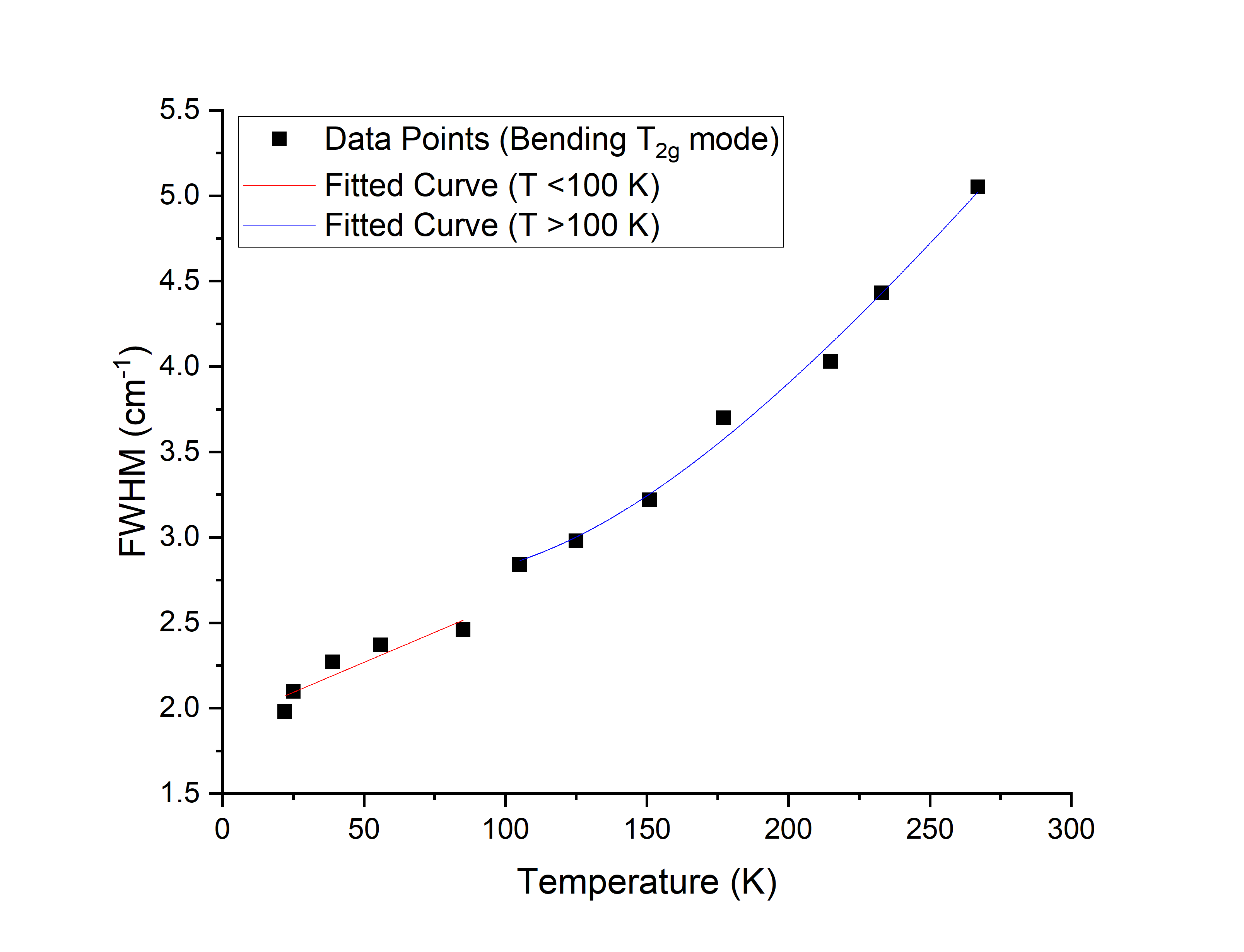}
    \caption{Bending T$_{2g}$: FWHM}
\end{subfigure}
\hfill
\begin{subfigure}{0.48\textwidth}
    \includegraphics[width=\textwidth]{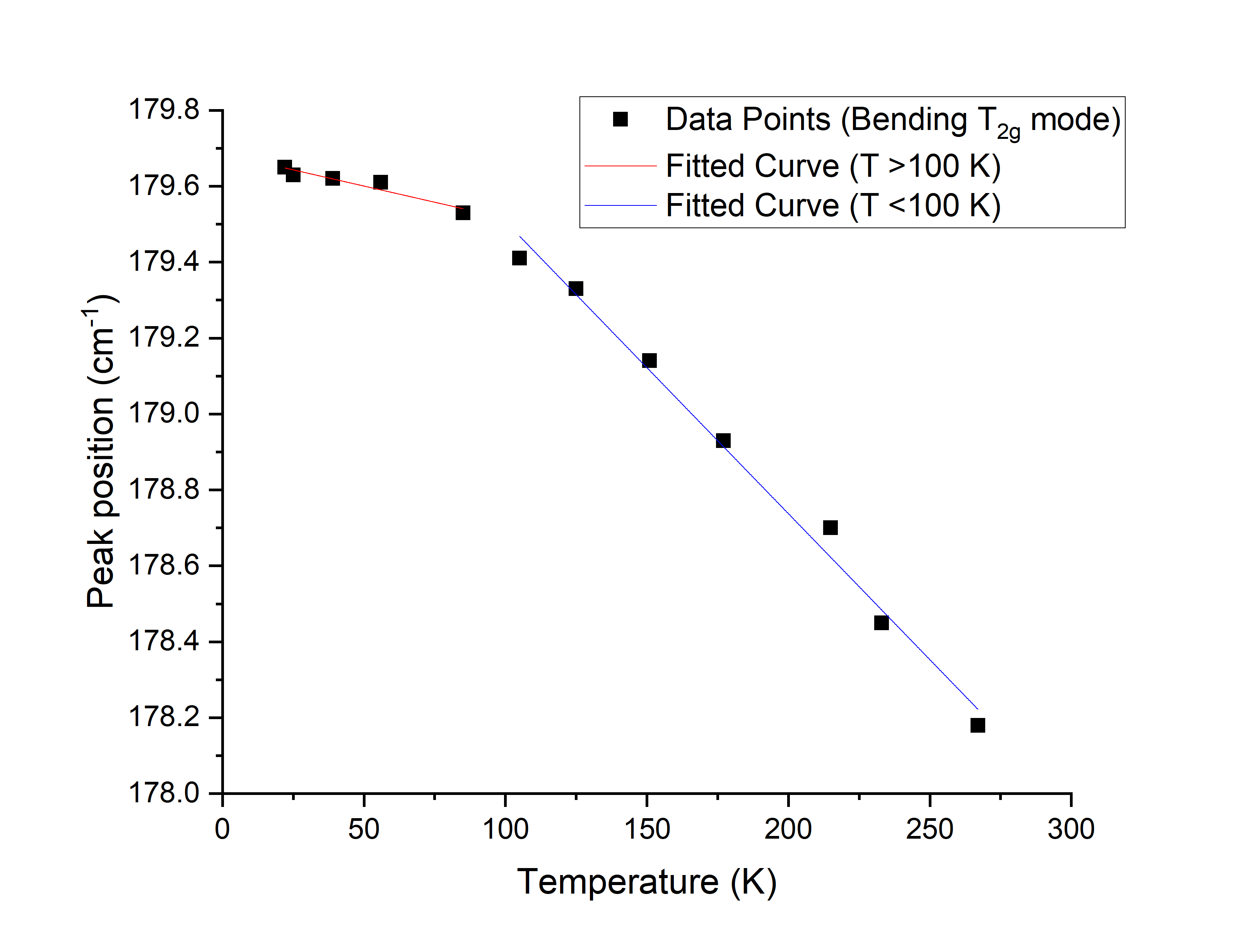}
    \caption{Bending T$_{2g}$: Position}
\end{subfigure}
     
\caption{Temperature dependence of FWHM and peak position for translational and bending T$_{2g}$ modes of Cs$_2$Ti$_{(1-x)}$Sb$_x$Cl$_6$ (x = 2\%). Solid lines represent fits to Eqs. (1) and (2).}  
\label{fig13a}
\end{figure}

\begin{figure}[htbp]
\centering
\begin{subfigure}{0.48\textwidth}
    \includegraphics[width=\textwidth]{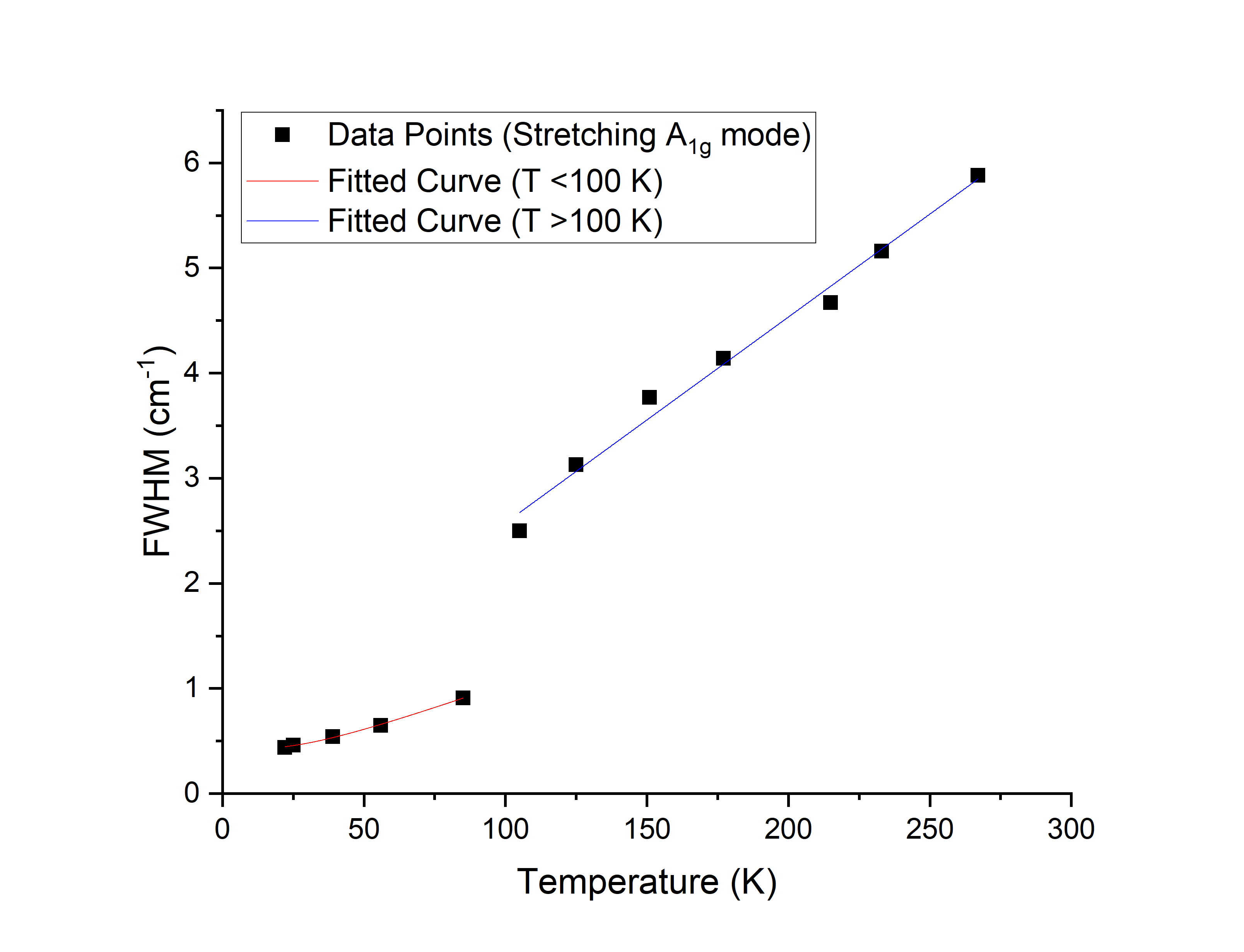}
    \caption{Stretching A$_{1g}$: FWHM}
\end{subfigure}
\hfill
\begin{subfigure}{0.48\textwidth}
    \includegraphics[width=\textwidth]{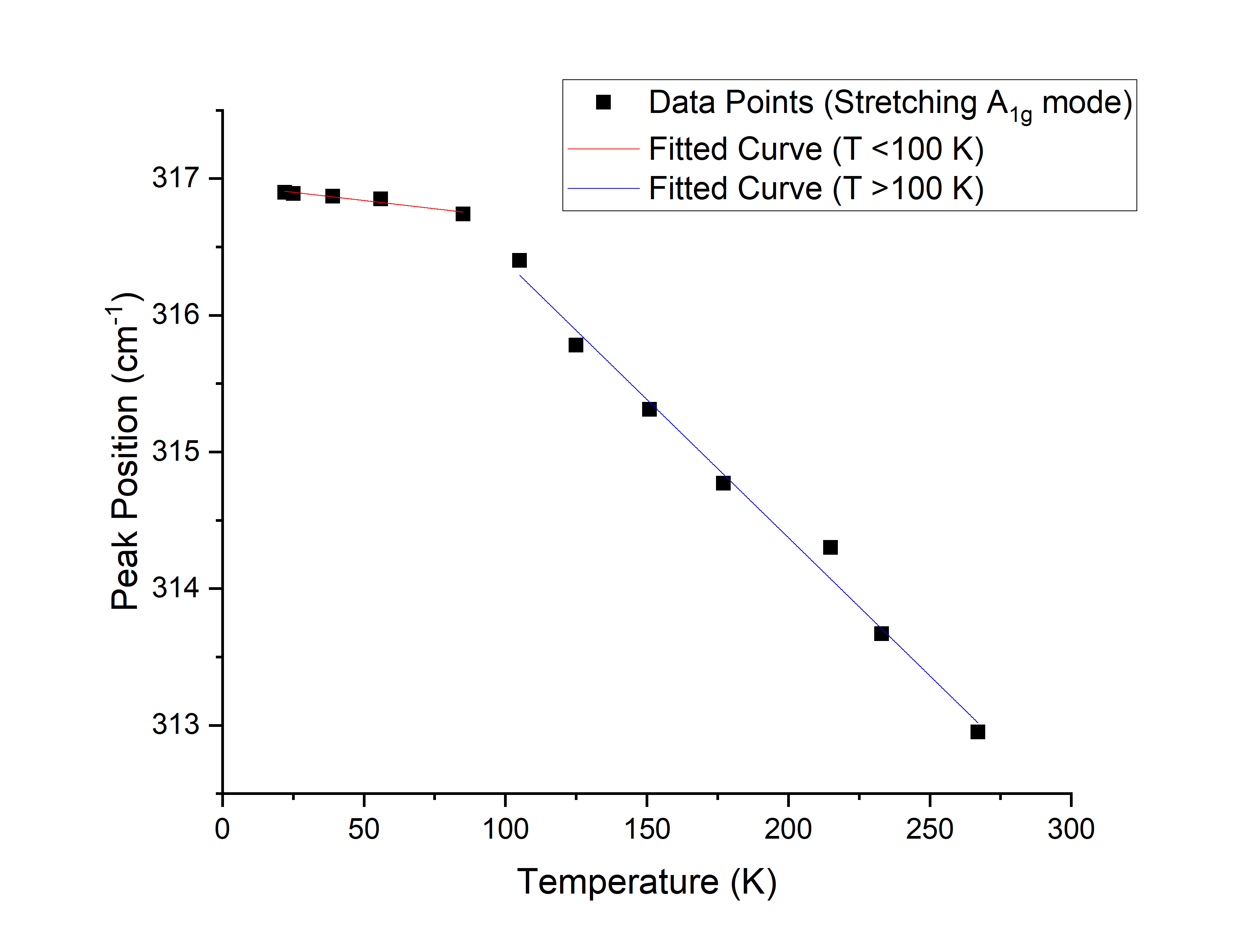}
    \caption{Stretching A$_{1g}$: Position}
\end{subfigure}

\vspace{0.3cm}

\begin{subfigure}{0.48\textwidth}
    \includegraphics[width=\textwidth]{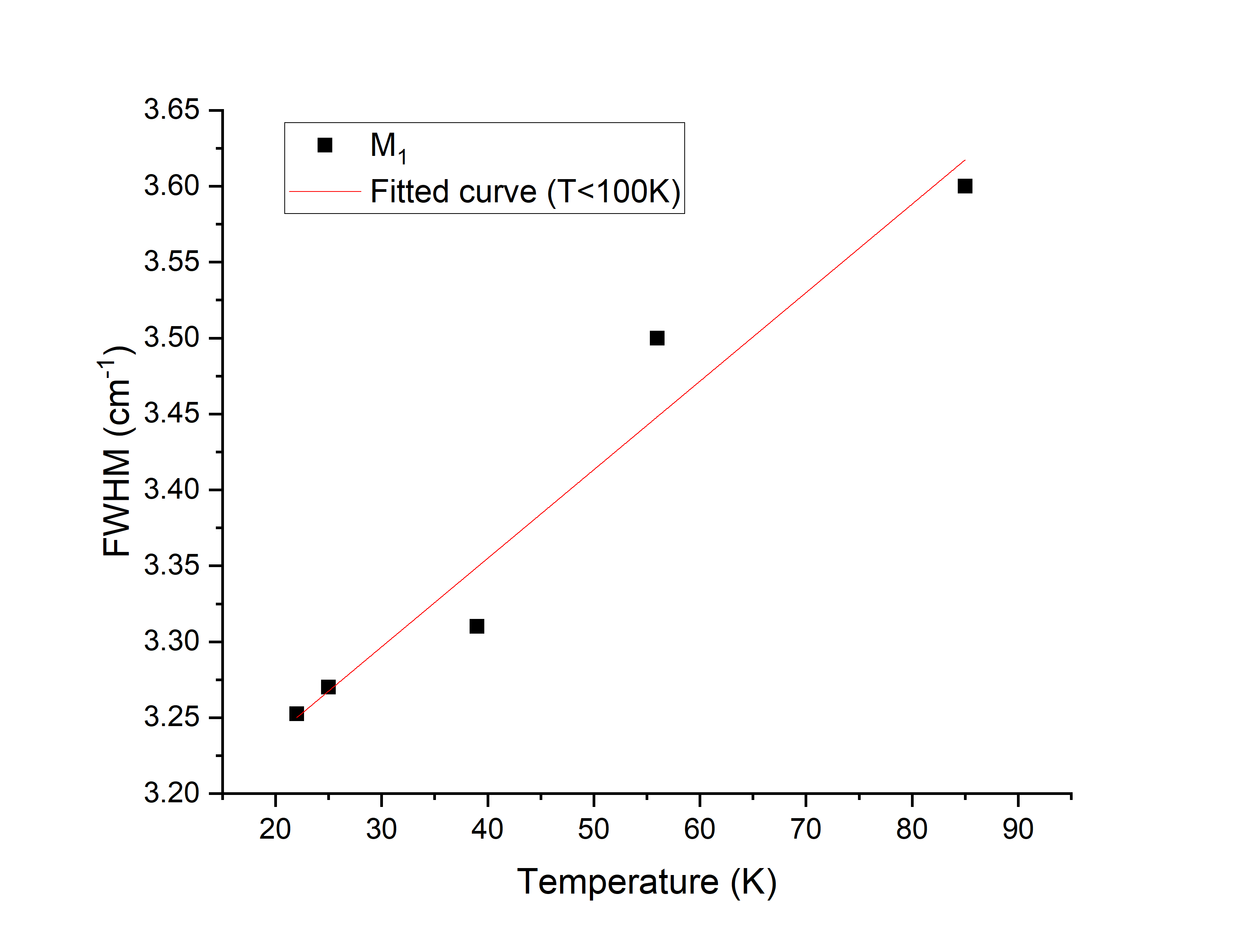}
    \caption{M$_1$ mode: FWHM}
\end{subfigure}
\hfill
\begin{subfigure}{0.48\textwidth}
    \includegraphics[width=\textwidth]{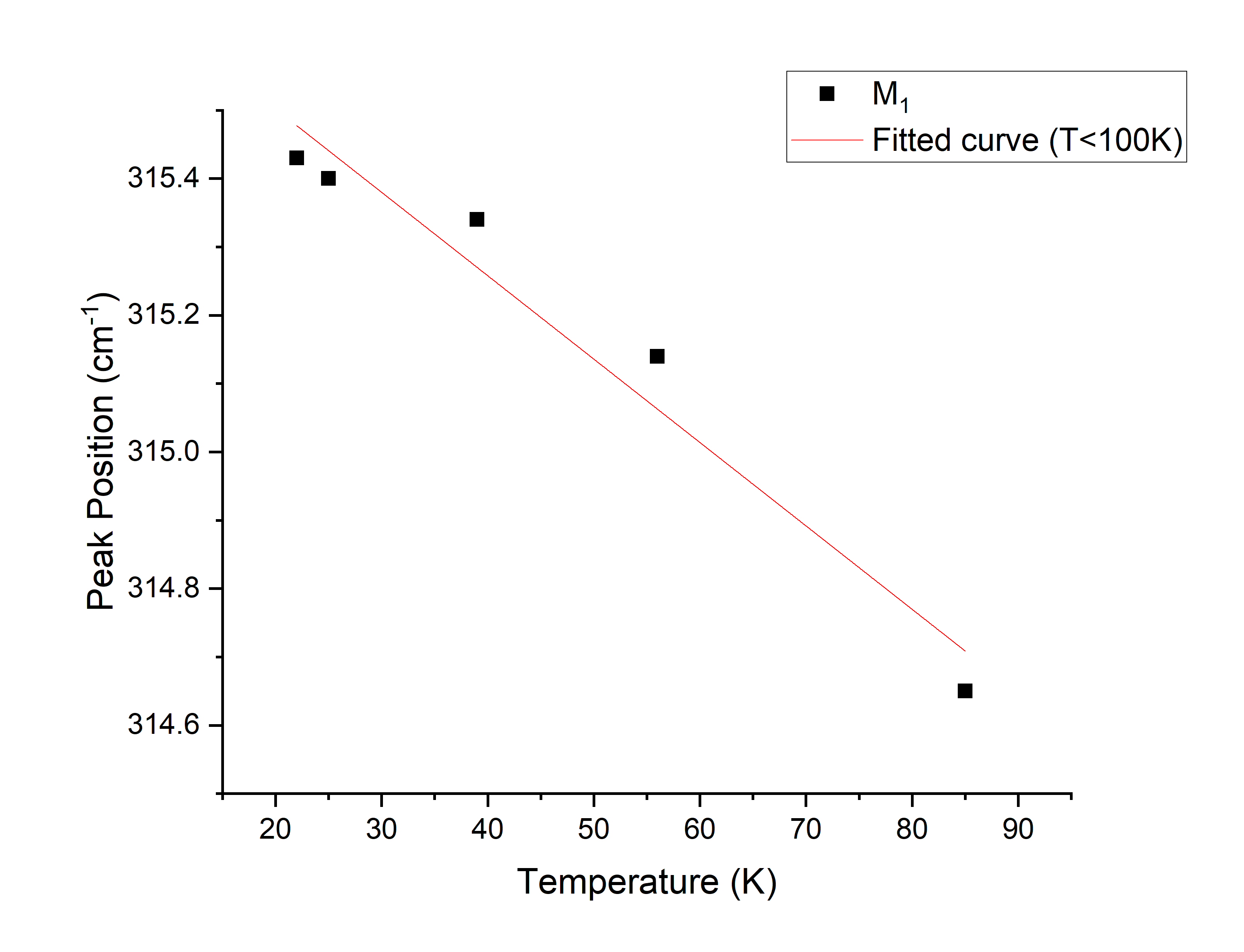}
    \caption{M$_1$ mode: Position}
\end{subfigure}
     
\caption{Temperature dependence of FWHM and peak position for stretching A$_{1g}$ and M$_1$ modes of Cs$_2$Ti$_{(1-x)}$Sb$_x$Cl$_6$ (x = 2\%). The M$_1$ mode appears only below 100 K. Solid lines represent fits to Eqs. (1) and (2).}  
\label{fig14}
\end{figure}

The temperature dependence of peak positions and FWHM were analyzed using the Grüneisen model and three-phonon anharmonic decay:
\begin{equation}
\omega (T) = \omega_0 + \chi T
\label{peakposition}
\end{equation}
\begin{equation}
\Gamma(T) = \Gamma(0) + A\left[1 + \frac{2}{e^{\hbar\omega/2k_BT}-1}\right]
\label{lifetime}
\end{equation}
where $\omega_0$ is the mode frequency at 0 K, $\chi$ is the first-order temperature coefficient, $\Gamma(0)$ is the intrinsic phonon linewidth, and $A$ is the anharmonic constant. All parameters were fitted separately for $T < 100$ K and $T > 100$ K regions. Tables \ref{undt}, \ref{inct}, and \ref{sbdt} summarize the fitted parameters for all three samples. The fitted modes are displayed in Figures \ref{Fig8}, \ref{Fig11a}, \ref{Fig11b}, \ref{fig13a}, and \ref{fig14}.

For pristine Cs$_2$TiCl$_6$, both $\chi$ and $A$ change across 100 K, with the FWHM broadening indicating optical phonon decay into two acoustic phonons. The Sb-containing samples show more pronounced changes in $\chi$ and $A$ across 100 K, accompanied by the emergence of the new M$_1$ mode below 100 K. 

While these observations could suggest a structural phase transition near 100 K in Sb-doped systems, several alternative explanations such as: (i) M$_1$ mode may be a Raman-inactive mode of the Fm$\bar{3}$m structure that becomes visible due to local symmetry breaking induced by Sb dopants, without requiring a global structural phase transition , or (ii) presence of secondary phase below 100 K must be considered. Low-temperature X-ray diffraction can be considered to gain insights of the new observed Raman mode.

\begin{table}[htbp]
\centering
\caption{Temperature dependence parameters for Raman modes of Cs$_2$TiCl$_6$}
\label{undt}
\begin{tabular}{lcccccc}
\hline\hline
 & $\omega(0)$ & $\Gamma(0)$ & $\chi$ ($T < 100$ K) & $\chi$ ($T > 100$ K) & A ($T < 100$ K) & A ($T > 100$ K) \\
Raman Mode & (cm$^{-1}$) & (cm$^{-1}$) & ($10^{-4}$ cm$^{-1}$/K) & ($10^{-4}$ cm$^{-1}$/K) & (cm$^{-1}$) & (cm$^{-1}$) \\
\hline
Bending T$_{2g}$ & 185.459$\pm$0.046 & 3.64 & $-29.2\pm8.2$ & $-46.7\pm2.5$ & 1.19 & 0.0085 \\
Stretching A$_{1g}$ & 322.565$\pm$0.056 & 3.93 & $-66.9\pm9.9$ & $-154.3\pm4.3$ & 0.0003 & 1.27 \\
\hline\hline
\end{tabular}
\end{table}

\begin{table}[htbp]
\centering
\caption{Temperature dependence parameters for Raman modes of Cs$_2$TiCl$_6$ (3\% Sb)}
\label{inct}
\begin{tabular}{lcccccc}
\hline\hline
 & $\omega(0)$ & $\Gamma(0)$ & $\chi$ ($T < 100$ K) & $\chi$ ($T > 100$ K) & A ($T < 100$ K) & A ($T > 100$ K) \\
Raman Mode & (cm$^{-1}$) & (cm$^{-1}$) & ($10^{-4}$ cm$^{-1}$/K) & ($10^{-4}$ cm$^{-1}$/K) & (cm$^{-1}$) & (cm$^{-1}$) \\
\hline
Translational T$_{2g}$ & 54.515$\pm$0.016 & 0.0228 & $-83.5\pm9.7$ & $-100.7\pm2.6$ & 0.00013 & $-0.019$ \\
Bending T$_{2g}$ & 183.405$\pm$0.016 & 1.73 & $-52.7\pm3.2$ & $-29.3\pm3.1$ & 6.006 & 6.22 \\
Stretching A$_{1g}$ & 320.67$\pm$0.026 & 0.0058 & $-47.3\pm5.0$ & $-103.6\pm5.28$ & 0.384 & 4.036 \\
M$_1$ ($T < 100$ K) & 319.215$\pm$0.052 & 1.239 & $-100\pm10$ & --- & $-0.0027$ & --- \\
\hline\hline
\end{tabular}
\end{table}

\begin{table}[htbp]
\centering
\caption{Temperature dependence parameters for Raman modes of Cs$_2$TiCl$_6$ (2\% Sb)}
\label{sbdt}
\begin{tabular}{lcccccc}
\hline\hline
 & $\omega(0)$ & $\Gamma(0)$ & $\chi$ ($T < 100$ K) & $\chi$ ($T > 100$ K) & A ($T < 100$ K) & A ($T > 100$ K) \\
Raman Mode & (cm$^{-1}$) & (cm$^{-1}$) & ($10^{-4}$ cm$^{-1}$/K) & ($10^{-4}$ cm$^{-1}$/K) & (cm$^{-1}$) & (cm$^{-1}$) \\
\hline
Translational T$_{2g}$ & 50.646$\pm$0.015 & 0.729 & $-49.9\pm3.01$ & $-123.4\pm6.8$ & 0.0008 & 2.21 \\
Bending T$_{2g}$ & 179.685$\pm$0.014 & 1.916 & $-17.3\pm2.9$ & $-76.9\pm3.6$ & $-0.0004$ & 4.92 \\
Stretching A$_{1g}$ & 316.960$\pm$0.018 & 0.0058 & $-24.3\pm3.7$ & $-202.3\pm9.23$ & 0.423 & 0.013 \\
M$_1$ ($T < 100$ K) & 315.746$\pm$0.013 & 3.122 & $-122.6\pm15.1$ & --- & 0.0011 & --- \\
\hline\hline
\end{tabular}
\end{table}

\subsection{High-Pressure Raman and Photoluminescence}

High-pressure Raman measurements were performed on Cs$_2$TiCl$_6$ (2\% Sb) up to 30 GPa using a diamond anvil cell (DAC). Samples were loaded in a 290 $\mu$m steel gasket with a 4:1 methanol:ethanol mixture as pressure-transmitting medium to maintain hydrostatic conditions. Pressure was calibrated using the ruby fluorescence R$_2$ line shift \cite{rubyf}:
\begin{equation}
P = \frac{A}{B}\left[\left(\frac{\lambda}{\lambda_0}\right)^B - 1\right]
\end{equation}
where $\lambda_0$ = 694.236 nm, A = 1904 GPa, and B = 7.665. Raman spectra were collected using 532 nm excitation (power $\leq$ 3.5 mW, 1500 lines/mm grating). High-pressure photoluminescence was measured at selected pressures using 488 nm excitation over 500-850 nm.

\begin{figure}[htbp]
\centering
 \includegraphics[width=0.7\textwidth]{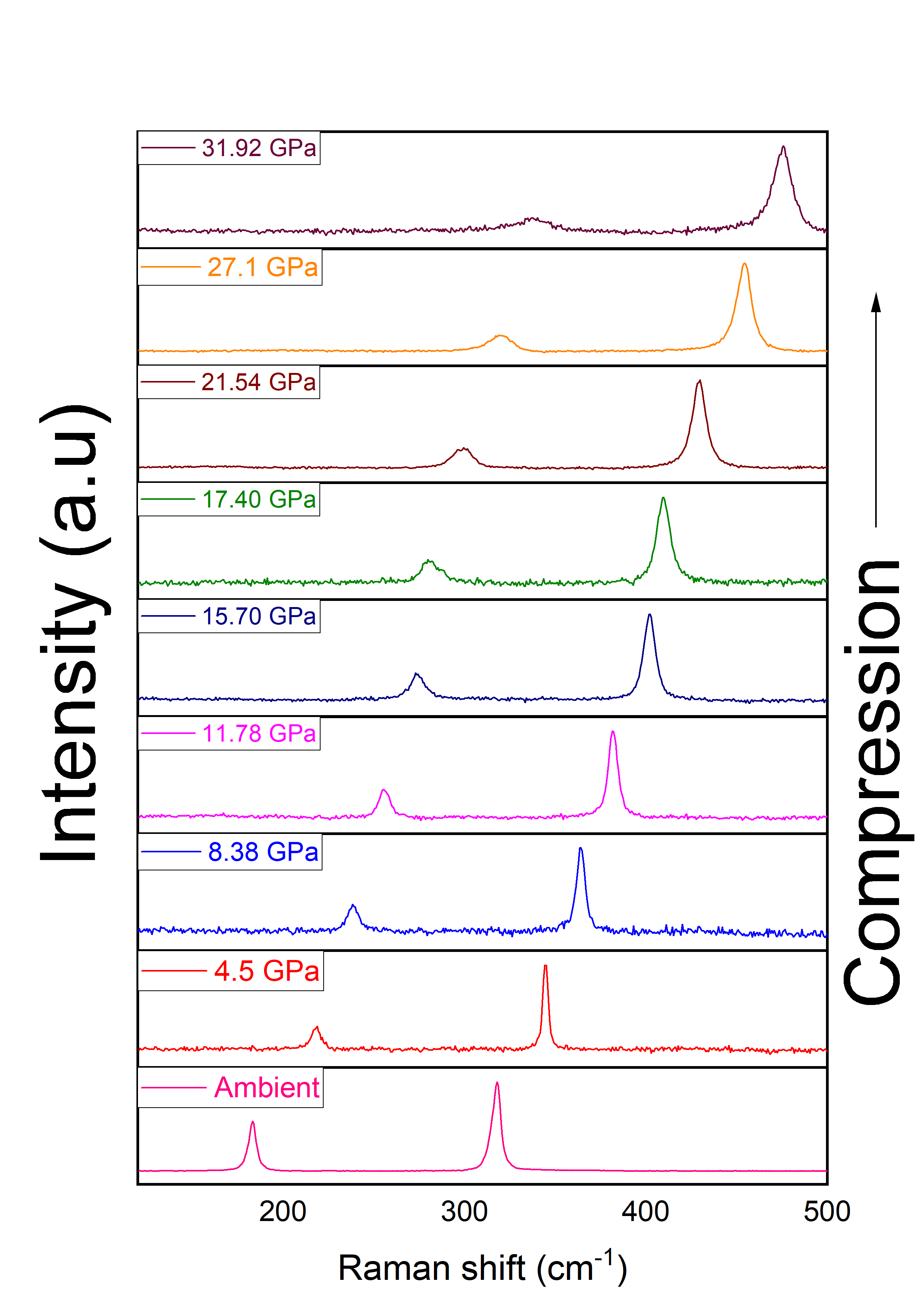}   
 \caption{Pressure-dependent Raman spectra of Cs$_2$Ti$_{(1-x)}$Sb$_x$Cl$_6$ (x = 2\%) from ambient to 30 GPa}
 \label{Fig15}
\end{figure}

\begin{figure}[htbp]
\centering
\begin{subfigure}{0.48\textwidth}
    \includegraphics[width=\textwidth]{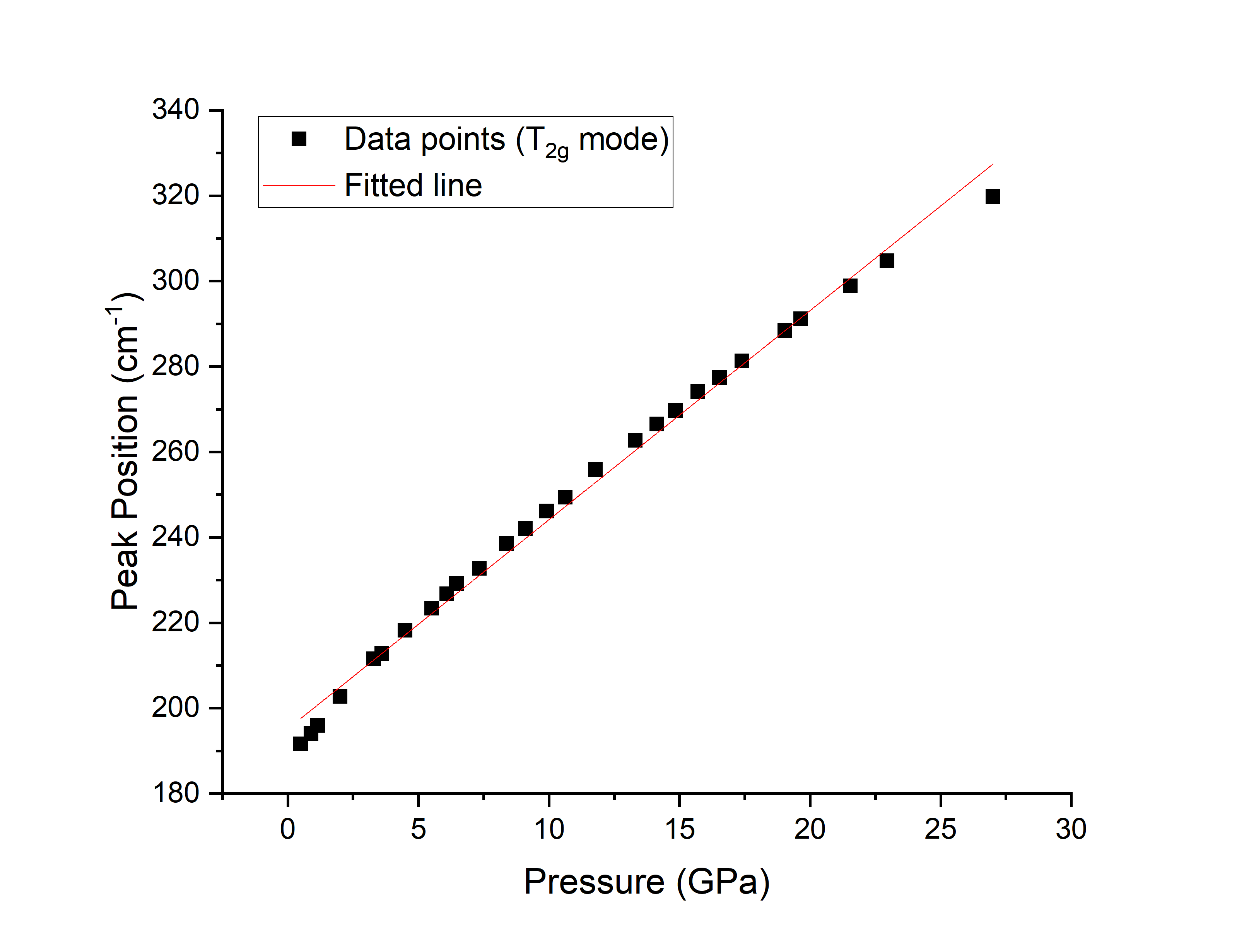}
    \caption{T$_{2g}$ mode}
\end{subfigure}
\hfill
\begin{subfigure}{0.48\textwidth}
    \includegraphics[width=\textwidth]{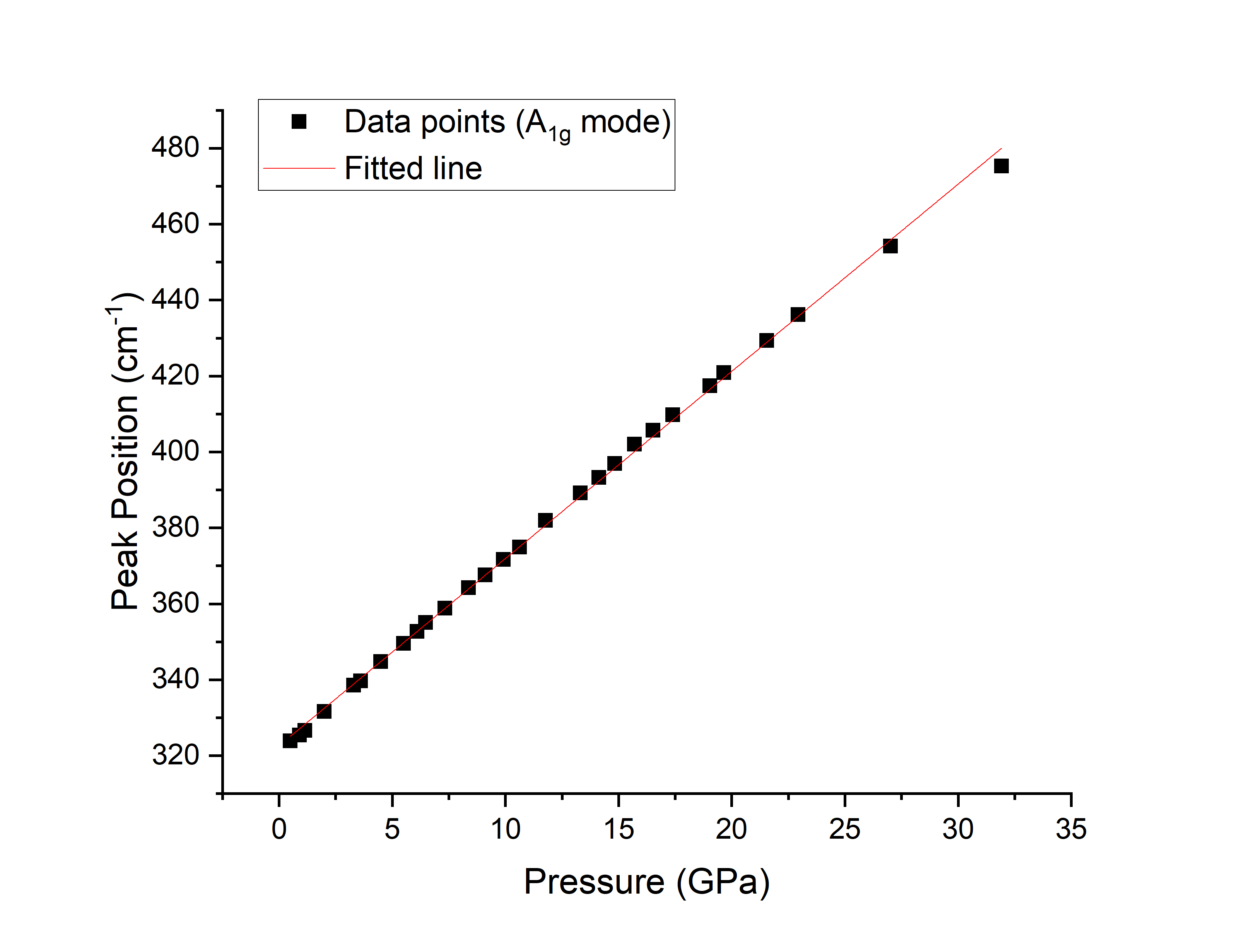}
    \caption{A$_{1g}$ mode}
\end{subfigure}     
\caption{Pressure dependence of Raman mode frequencies in Cs$_2$Ti$_{(1-x)}$Sb$_x$Cl$_6$ (x = 2\%)}  
\label{Fig16}
\end{figure}

\begin{figure}[htbp]
\centering
 \includegraphics[width=0.7\textwidth]{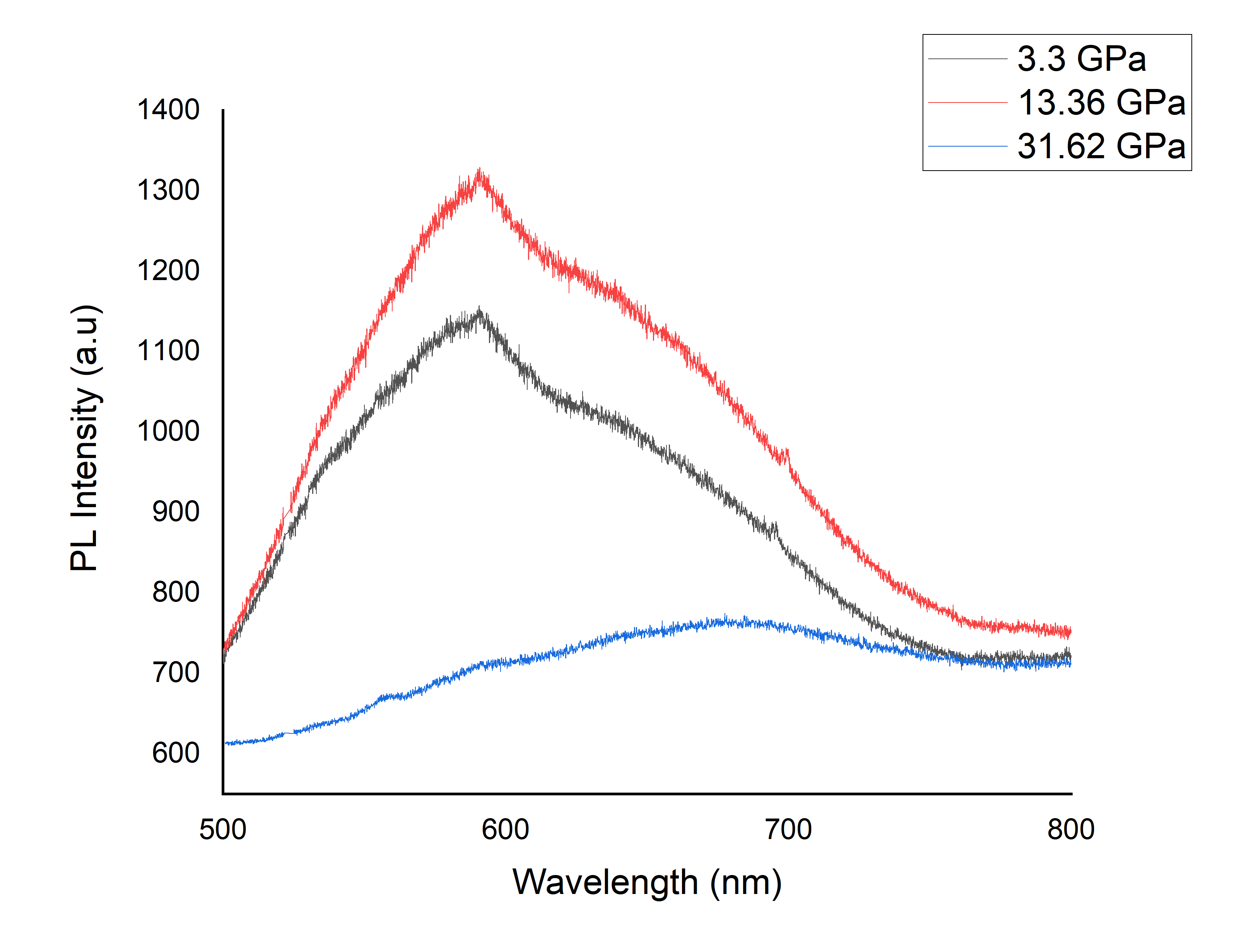}   
 \caption{Pressure-dependent photoluminescence of Cs$_2$Ti$_{(1-x)}$Sb$_x$Cl$_6$ (x = 2\%) at three selected pressures}
 \label{Fig17}
\end{figure}

Figure \ref{Fig15} shows the pressure evolution of Raman spectra up to 30 GPa. Both T$_{2g}$ and A$_{1g}$ modes shift continuously to higher frequencies with increasing pressure (Figure \ref{Fig16}), with similar pressure coefficients throughout the measured range. The FWHM shows no systematic pressure dependence. The continuous evolution of Raman modes without discontinuities suggests no major structural phase transitions occur up to 30 GPa, though subtle transitions could be missed by Raman spectroscopy alone. Photoluminescence spectra (Figure \ref{Fig17}) redshift with increasing pressure, consistent with pressure-induced modification of self-trapped exciton states. The lattice compression under pressure alters the carrier-phonon coupling strength, leading to the observed spectral evolution.

\section{Conclusion}

We have synthesized vacancy-ordered double perovskite Cs$_2$TiCl$_6$ and systematically investigated the effects of Sb and Bi doping through structural, vibrational, optical, and stability characterization. Powder XRD confirms the cubic Fm$\bar{3}$m structure with reduced impurity peak intensity in Sb-doped samples compared to pristine and Bi-doped variants, though all samples contain residual impurity phases that complicate quantitative interpretation. Room temperature Raman spectroscopy reveals three main phonon modes corresponding to [TiCl$_6$]$^{2-}$ octahedral vibrations, with Sb-doped samples exhibiting cleaner spectra with fewer impurity-related peaks.

Low-temperature Raman measurements (4 to 273 K) reveal systematic hardening and sharpening of all phonon modes upon cooling, indicating reduced anharmonicity and increased phonon lifetimes. Both the first-order temperature coefficient $\chi$ and anharmonic constant $A$ exhibit changes across 100 K in all samples, more pronounced in Sb-doped materials. Most notably, Sb-doped samples (both 2\% and 3\%) display an additional mode M$_1$ appearing exclusively below 100 K near 314--319 cm$^{-1}$, which is absent in pristine Cs$_2$TiCl$_6$. While this anomalous behavior could indicate a structural phase transition, multiple alternative explanations are equally plausible, including disorder-activated Raman modes, dopant-induced local symmetry breaking, combination modes, or minor secondary phases. The proximity of M$_1$ to the A$_{1g}$ mode frequency and the presence of residual impurities further complicate interpretation. Low-temperature X-ray diffraction measurements are essential to definitively establish dopant induced structural phase transition at low temperature. 

Photoluminescence measurements show broad emission centered at 448 nm in all samples, characteristic of self-trapped exciton (STE) formation due to strong exciton-phonon coupling within isolated [TiCl$_6$]$^{2-}$ octahedra. Sb-doped samples exhibit larger emission FWHM (164.73 nm) compared to Bi-doped samples (138.2 nm), which may indicate enhanced electron-phonon coupling or structural inhomogeneity. High-pressure Raman spectroscopy up to 30 GPa on Cs$_2$TiCl$_6$ (2\% Sb) reveals continuous mode hardening with no discontinuities, indicating structural stability across this pressure range. Pressure-dependent PL shows systematic redshifts consistent with pressure-modified STE states. Stability tests confirm thermal endurance up to 120$^{\circ}$C and selective water interaction that removes impurity-related Raman modes while preserving the core perovskite structure.

These results demonstrate that Sb doping modulates the vibrational and optical properties of Cs$_2$TiCl$_6$ and improves phase purity compared to pristine samples. The anomalous low-temperature behavior in Sb-doped samples represents an interesting observation that requires further investigation. However, the underlying mechanism remains unclear from the current data. Future work such as (i) low-temperature structural characterization (synchrotron XRD) to definitively identify any symmetry changes, (ii) compositional analysis to verify actual Sb incorporation, and (iii) theoretical calculations to predict possible structural modifications and their Raman signatures is required to interpret the observed low temperature behavior. Understanding the origin of the M$_1$ mode and the role of Sb doping in modulating structural and vibrational properties would provide valuable insights for developing lead-free halide perovskites for optoelectronic applications.

\section*{Acknowledgments}

I would like to express my sincere gratitude to my supervisor, Prof. Goutam Dev Mukherjee, for his constant guidance, support, and encouragement throughout this work. I am deeply grateful to Prof. Sayan Bhattacharyya for his valuable insights and for providing the platform to synthesize samples for this study. I extend my thanks to all members of the High Pressure Physics Laboratory for their assistance and support. I acknowledge IISER Kolkata for providing the research facilities and infrastructure that made this work possible. This work was carried out as part of Aalok Tiwari's Master's thesis project(18MS) at DPS, IISER Kolkata.

\bibliographystyle{ieeetr}
\bibliography{references}
\newpage
\section{Figures}

\end{document}